\documentclass[aps,prb,twocolumn,superscriptaddress,longbibliography,floatfix]{revtex4-2}
\usepackage{amsmath,amssymb,bm,graphicx,xcolor,booktabs}
\usepackage[colorlinks=true,citecolor=blue,urlcolor=blue,linkcolor=blue]{hyperref}

\AtBeginDocument{\raggedbottom}
\newcommand{\kk}{\mathbf k}

\newcommand{\Lcal}{\mathcal L}
\newcommand{\Mcal}{\mathcal M}

\newcommand{\tbg}{\tau_{\rm bg}}

\begin{document}
\title{Spin-conserving and spin-flip magnetic scattering in the altermagnetic spin-splitter effect}
\author{Yunxi Jiang}
\affiliation{Universit\'e de Strasbourg, CNRS, Institut de Physique et Chimie des
Mat\'eriaux de Strasbourg, UMR 7504, F-67000 Strasbourg, France}
\author{Hao Yu}
\email{Hao.Yu@xjtlu.edu.cn}
\affiliation{Department of Physics, Xi'an Jiaotong-Liverpool University, 215123, Suzhou, China}
\begin{abstract}
Critical magnetic fluctuations scatter electrons strongly near an ordering transition, but they barely degrade the charge current, because they scatter mostly through small angles. In an altermagnet the electric field also drives a spin current, and a nearly forward spin flip moves an electron to the other spin band at almost unchanged velocity. We develop a Boltzmann theory of the spin splitter ratio that keeps the full collision operator of quasistatic magnetic fluctuations and its change with the magnetic order. Fluctuations of equal total strength give the same resistivity for every fraction of spin flips, yet they can move the spin splitter ratio in opposite directions. Spin conserving fluctuations relax the spin current at about the charge relaxation rate, whereas forward flips equalize the two spin populations at equal energy at the quasiparticle rate, which in two dimensions grows linearly with the correlation length. Energy conservation on the split bands protects the part of the spin current carried by the band shift of the occupied states at every scattering strength, and the splitting blocks forward flips once the spin Fermi lines lie farther apart than the inverse correlation length. When flips dominate, a quarter to a half of the lattice band ratio survives. The transport time approximation reproduces the initial growth of the resistivity but recovers only a quarter to a half of the initial fall of the ratio, and a disorder series at fixed reduced temperature measures this fall independent of the order amplitude.
\end{abstract}
\maketitle

\section{Introduction}
In an altermagnet, collinear order with zero net magnetization splits the bands by spin, and the splitting changes sign with crystal momentum \cite{PhysRevX.12.040501,Hayami2019,Ahn2019,Zeng2026,Jin2024}. An electric field then drives a spin current along with the charge current, the spin splitter effect \cite{Naka2019,PhysRevLett.126.127701}.
For a \(d_{x^2-y^2}\) splitting, the spin conductivity tensor satisfies \(\sigma^s_{xx}=-\sigma^s_{yy}\) and \(\sigma^s_{xy}=0\). An electric field applied along a crystal diagonal therefore generates a transverse spin current. The component \(\sigma^s_{xx}\) describes the same response in the coordinate system aligned with the crystal axes.

Near the ordering temperature $T_N$, the strongest scatterers are the critical fluctuations of the order itself. They broaden the electronic states, but the charge current survives, because their long wavelength part scatters mostly through small angles. Fisher and Langer showed that short distance spin correlations govern the magnetic resistivity \cite{Fisher1968}, and the quasiparticle rate and the transport rate can grow differently \cite{DellAnnaMetzner2006,DellAnnaMetzner2007,WangBerg2019}. This raises the question of whether the spin current survives as well. An altermagnet has two spin bands with equal densities of states and different Fermi lines, so a nearly forward spin flip moves an electron between them at almost unchanged velocity.

Some calculations of the spin splitter effect use a constant relaxation time for all electronic states, which multiplies the spin and charge conductivities by the same factor and fixes the spin splitter ratio $\theta$ from the bands alone \cite{Dou2025}.
First principles calculations show that scattering by phonons can enhance, suppress or reverse the spin splitter effect \cite{He2025} and enhances its response at finite frequency \cite{Hodt2026}. Spin dependent quasiparticle lifetimes have been calculated \cite{Leraand2026}, and the errors of the self energy and momentum relaxation time approximations for charge mobilities have been assessed \cite{Ponce2020,Claes2022}. Transport studies near Pomeranchuk and Ising nematic transitions established that the lifetime of a momentum state and the decay of a current are different quantities \cite{DellAnnaMetzner2006,DellAnnaMetzner2007,WangBerg2019,MaslovYudsonChubukov2011,Hartnoll2014}. Mesoscale theory addresses textures and spin pumping \cite{Zarzuela2025}. Here we ask how scattering into a state, and the change of the accessible final states with the order, act on the spin splitter current when the scatterers are the critical fluctuations of the order itself.

We develop a Boltzmann theory that explicitly resolves the momentum and spin of the final states in scattering by quasistatic magnetic fluctuations. The theory also accounts for how magnetic order changes the final states accessible to scattering.
The response is linear in the electric field and arises from the collinear magnetic order alone, with spin orbit coupling neglected. We find the following results.
(i) The slope of $\theta$ at weak scattering compares the initial decay rates of the spin and charge currents, and it interpolates linearly in the spin flip fraction between a spin conserving and a spin flip limit, so fluctuations with the same resistivity can move $\theta$ in opposite directions. (ii) Each spin drifts with its own velocity, so the field leaves the two spin populations unequal at equal energy. Nearly forward flips remove this imbalance at the quasiparticle rate, so in two dimensions the spin flip slope grows nearly linearly with the correlation length. At every scattering strength, energy conservation on the split bands protects the part carried by the band shift of the occupation, and the splitting blocks forward flips once the correlation length exceeds the inverse separation of the two spin Fermi lines. (iii) On the nearest neighbor square lattice band, spin flips drive the reduced ratio to the velocity only value, which in the momentum partition is the cancellation of the occupation contribution. In a parabolic band, where a constant relaxation time gives exactly the band ratio by a Drude argument, every deviation of the ratio comes from scattering. (iv) When spin flips limit both currents, the ratio settles on a plateau of a quarter to a half of the lattice band ratio, which approaches the spin current of the charge carrying distribution placed on the two split Fermi lines as forward flips dominate. (v) The transport time approximation reproduces the initial growth of the resistivity exactly but recovers only a quarter to a half of the initial fall of $\theta$, and a disorder series at fixed reduced temperature measures this fall independent of the order amplitude.

This Article is the companion of the Letter \cite{JiangLetter2026}. The Letter reports the central result, that the spin splitter current shares the protection of the charge current against critical scattering only when the fluctuations conserve spin, and proposes the disorder series as its test. Here we derive the collision operator and its change with the order, prove the forward flip estimate and the exact constant time ratio of the parabolic band, resolve the spin current into velocity and occupation contributions, compare the full theory with relaxation time approximations, and treat finite splitting, strong spin flips and the conditions of the disorder series in full.

\section{Spin split bands and spin dependent magnetic scattering}\label{sec:model}
\subsection{Spin split bands and the two contributions to the spin current}
We consider a single electronic band with momentum $\kk$, unsplit dispersion $\epsilon_0(\kk)$ and chemical potential $\mu$. The magnetic order enters through the dimensionless amplitude $\phi_0$, and the function $d_{\mathbf k}$, with units of energy, fixes the momentum dependence of the spin splitting. The spin label $\sigma=\pm1$ refers to a fixed axis $z$, and reversal of the magnetic domain corresponds to $\phi_0\to-\phi_0$. The two spin bands are
\begin{equation}
\varepsilon_{\kk\sigma}=\epsilon_0(\kk)-\mu+\sigma\phi_0d_\kk,
\label{eq:band}
\end{equation}
and their energies differ by $2|\phi_0d_{\mathbf k}|$. Along Cartesian direction $i$, the velocity of the unsplit band is $v_i$, and $u_i$ is its change per unit $\sigma\phi_0$,
\begin{equation}
v_i=\frac{1}{\hbar}\partial_{k_i}\epsilon_0,\qquad
u_i=\frac{1}{\hbar}\partial_{k_i}d_\kk,\qquad
v_{i,\sigma}=v_i+\sigma\phi_0u_i.
\label{eq:velocities}
\end{equation}

We keep the coefficient linear in $\phi_0$ at fixed electron density, temperature, correlation length and magnetic scattering strengths. The fluctuation strengths are even under $\phi_0\to-\phi_0$, so keeping them fixed is exact to linear order. When longitudinal and transverse magnetic fluctuations are uncorrelated, the Born collision operator preserves the subspace spanned by $I$ and $\sigma_z$, and the response to the electric field is described by the occupations of the two spin states. An antisymmetric chiral component of the transverse correlator also preserves this subspace and makes the two opposite spin flip processes unequal at the same electron energy. For classical static fluctuations in a centrosymmetric magnet, complex conjugation together with spatial inversion makes the static correlator real and symmetric in spin space, so this component vanishes, and its quantum part is smaller by a factor $\hbar\omega_q/k_BT$ and disappears in the quasistatic limit (Supplemental Material Sec.~S1). Altermagnetic order splits magnons of opposite chirality with a $d$-wave dependence, so this quantum part has the symmetry of the response computed here. It drops out only because $\hbar\omega_q/k_BT$ is small, which holds near $T_N$, where the fluctuations slow down.

Let $\delta f_\sigma(\mathbf k)$ be the change in the occupation of momentum $\mathbf k$ produced by the electric field. Its average $\delta f^+=(\delta f_\uparrow+\delta f_\downarrow)/2$ and half difference $\delta f^-=(\delta f_\uparrow-\delta f_\downarrow)/2$ describe the common motion and the relative motion of the two spins. The electron charge is $-e$ with $e>0$. We express the angular momentum current $J_i^{S_z}$ in electrical units as $J_i^s=-(2e/\hbar)J_i^{S_z}$, and the momentum integral $\int_{\mathbf k}$ counts one state per spin per unit area, so that charge and spin conductivities have the same units. For a field $E_j$, the currents to first order in $\phi_0$ are
\begin{align}
J_i^s&=-2e\int_\kk\left[\phi_0u_i\delta f^++v_i\delta f^-\right],\nonumber\\
J_j^c&=-2e\int_\kk v_j\delta f^+,
\label{eq:currents}
\end{align}
which define the conductivities through $J_i^s=\sigma^s_{ij}E_j$ and $J_j^c=\sigma^c_{jj}E_j$.

The first term of $J_i^s$ is the velocity contribution, the common occupation weighted by the velocity difference of the two spins. The second is the occupation contribution, the occupation difference weighted by the ordinary velocity. In the uniform bulk, inversion symmetry gives $\int_{\mathbf k}\delta f^-=0$, while $\int_{\mathbf k}v_i\delta f^-$ can be finite, so the occupation contribution carries spin with zero net spin accumulation. This split compares the two spins at the same crystal momentum $\mathbf k$, and we call it the momentum partition. A comparison at the same energy, the energy partition, divides the same total differently, because the lines of constant energy of the two spins move with $\phi_0$. The two contributions depend on the partition, and their sum, the observable spin response, is the same in both.

In some band models the velocity change caused by the splitting is proportional to the ordinary velocity,
\begin{equation}
u_i=bv_i,
\label{eq:bandratio}
\end{equation}
with a constant band ratio $b$. The velocity contribution $V$ and the charge coefficient $C$, defined in Eq.~\eqref{eq:responses} below, then obey $V=bC$ exactly. With $S=V+P$ the total spin response, $P$ the occupation contribution and $Q=S/C$ the reduced spin splitter ratio, $Q=b+P/C$. Once $b$ is known, $Q-b$ measures the occupation contribution directly.

On the two dimensional square lattice, let $t_1>0$ be the hopping between nearest neighbors and $t_2$ the hopping between next nearest neighbors. With unit lattice constant and exchange scale $d_0=t_1$, the unsplit band and the spin splitting are
\begin{align}
\epsilon_0(\kk)&=-2t_1(\cos k_x+\cos k_y)-4t_2\cos k_x\cos k_y,\nonumber\\
d_\kk&=d_0(\cos k_x-\cos k_y),\qquad d_0=t_1.
\label{eq:latticeband}
\end{align}
The factor $\cos k_x-\cos k_y$ gives the splitting opposite signs along the $x$ and $y$ crystal axes, the defining feature of $d_{x^2-y^2}$ altermagnetism, and $\Phi_d=d_\kk/d_0$ is the dimensionless form factor. We use two reference bands, the band with only nearest neighbor hopping (NN band), $t_2=0$, and the band that also includes next nearest neighbor hopping (NNN band), $t_2/t_1=-0.2$. For the NN band, Eq.~\eqref{eq:bandratio} holds with $b=-d_0/(2t_1)=-1/2$.

For an electron scattered from momentum $\kk$ to $\kk'$, we measure the momentum transfer on the lattice by
\begin{equation}
q_{\rm lat}^2=4\sum_{\alpha=x,y}\sin^2[(k_\alpha-k'_\alpha)/2],
\label{eq:qlat}
\end{equation}
so that small $q_{\rm lat}$ means nearly forward scattering. We compare two forms of the coupling between the electrons and the magnetic fluctuations,
\begin{align}
\Lambda_A(\kk,\kk')&=[\Phi_d(\kk)+\Phi_d(\kk')]/2,\nonumber\\
\Lambda_G(\kk,\kk')&=\Lambda_A(\kk,\kk')e^{-q_{\rm lat}^2/8}.
\label{eq:microvertices}
\end{align}
The first averages the $d$-wave form factor of the initial and final states, and the factor $e^{-q_{\rm lat}^2/8}$ in $\Lambda_G$ suppresses scattering with large momentum transfer. Both couplings agree at zero momentum transfer and respect the periodicity of lattice momentum. At the same reference scattering strength, comparing them shows how much scattering through large angles matters. We quote numbers for $\Lambda_A$ unless stated otherwise, and $\Lambda_G$ gives the same trends.

As a continuum reference we use the parabolic band $\epsilon_0=\hbar^2k^2/(2m)$ with $d_{\mathbf k}=d_0(k_x^2-k_y^2)/k_F^2$, in units $m=d_0=\mu(0)=\hbar=1$, with $k_BT/\mu(0)=0.05$ and $\xi k_F=4$. For this band, Eq.~\eqref{eq:bandratio} holds with $b=d_0/\mu(0)$.

\subsection{Magnetic fluctuations and scattering strengths}
The magnetic fluctuations couple to the electrons through the exchange form factor of the band. A longitudinal fluctuation preserves the spin projection along $z$, and a transverse fluctuation connects opposite projections. Both kinds of fluctuations have the same magnetic correlation length $\xi$, the distance over which the fluctuating moments stay correlated, which sets the typical momentum transfer $1/\xi$. We insert the equal time equilibrium correlations of the fluctuations into an elastic collision integral. The fluctuations stay in equilibrium under the applied field, and the momentum they receive from the electrons passes on to the lattice. This defines the quasistatic limit of the transport problem \cite{MaslovYudsonChubukov2011,Hartnoll2014}. The elastic treatment requires the energy of the fluctuations that carry a given momentum transfer to be small compared with the temperature. 
Treating the magnetic fluctuations as an equilibrium bath assumes that their momentum relaxes faster than electrons transfer momentum to them. Magnon drag \cite{Sourounis2025} is beyond the scope of this approximation.
Transverse fluctuations of the N\'eel vector also come with fluctuations of the uniform magnetization, which couple through a constant form factor, and spin conserving two magnon processes add to the longitudinal scattering \cite{Maeland2024}.

The three spin components share the electron coupling because they come from a single vector exchange coupling. We give the longitudinal and transverse fluctuations the same momentum dependence and correlation length, with independent strengths $A_L$ and $A_T$. With this choice, the charge current depends only on the total strength $A_L+A_T$ (Sec.~\ref{sec:weak}). The probability of scattering with momentum transfer $q_{\rm lat}$ is proportional to $|\Lambda_m|^2/(\xi^{-2}+q_{\rm lat}^2)$ with $m=A,G$, and elastic energy conservation is imposed on the actual spin bands.

The background time $\tau_{\rm bg}$ describes momentum relaxation by defects and other sources unrelated to the magnetic fluctuations, and we take it independent of energy. For spin conserving pointlike impurities, the background rate is proportional to the density of states and depends on energy. Such a background changes the reference spin splitter ratio only through the thermal average, by less than one percent on the reference bands away from the van Hove saddle \cite{JiangLetter2026}.

On an elastic energy contour, let $w_b$ be the quadrature weight of state $b$, which contains the density of states, and let $W_{bc}$ be the rate from state $b$ to state $c$, including the final state measure. Detailed balance gives $w_bW_{bc}=w_cW_{cb}$. At contour energy $E$ we write $\delta f_b=F(E)\psi_b$, where $F(E)=-\partial f(E)/\partial E$ and $f$ is the Fermi function. The operator $\mathcal L$ relaxes $\psi$ with respect to the inner product $\langle A,B\rangle_w=\sum_b w_bA_bB_b$,
\begin{align}
(\Lcal\psi)_b&=\sum_c W_{bc}(\psi_b-\psi_c),\nonumber\\
\langle\psi,\Lcal\psi\rangle_w
&=\frac12\sum_{bc}w_bW_{bc}(\psi_b-\psi_c)^2.
\label{eq:collision}
\end{align}
This form keeps scattering both into and out of each state \cite{PhysRev.124.287,PhysRev.127.1391}. Summing the collision term with the weights $w_b$ gives zero, so scattering conserves electron number. The quadratic form is a sum of squared differences with positive rates, so it is nonnegative, and it vanishes only when $\psi$ is constant on each set of states connected by scattering.

The dimensionless strengths $K$ and $a$ stay fixed as $\phi_0$ varies. $K$ is twice the longitudinal charge decay rate on the unsplit Fermi line multiplied by $\tau_{\rm bg}$, and $a$ is twice the transverse quasiparticle scattering rate at the point of largest $|d_{\mathbf k}|$, multiplied by the same time. Because a flip changes the spin imbalance by two, $a$ is also $\tau_{\rm bg}$ times the relaxation rate of a uniform spin imbalance at that point. A purely longitudinal bath gives $x\simeq K/2$ at weak scattering. The relative weight of the two kinds of scattering is $R=A_T/A_L$, and equal fluctuations of the three spin components give $R=2$. The calibrated strengths obey $R=f_m a/K$, with $f_m=0.150$ for $\Lambda_A$ on the NN band, so $a=33$ at $K=0.1$ corresponds to $R\simeq49$ (Supplemental Material Sec.~S2). At $R=2$, $a/K=2/f_m$ grows nearly linearly with $\xi$, so a growing transverse rate coexists with a bounded transport strength.

We keep the electron density at its value for zero temperature and zero splitting, with $\mu(0)/t_1=-1$ for the NN band and $-1.2$ for the NNN band, and solve the chemical potential at every temperature and splitting. Unless it is varied explicitly, the correlation length is fixed by $\xi\bar k_F=4$, where
\begin{equation}
\bar k_F=\frac{\oint d\ell\,|\kk|/|\mathbf v_F|}{\oint d\ell/|\mathbf v_F|}
\label{eq:kf}
\end{equation}
is the mean radius of the unsplit Fermi line. On the NN band this gives $\xi\simeq2$ lattice spacings. Energies are in units of $t_1$ and rates in units of $\tau_{\rm bg}^{-1}$, and a physical realization requires $\hbar\Gamma_{\rm out}\ll E_F$ for the total rate $\Gamma_{\rm out}$ of scattering out of the states that carry the current. The Born approximation for fluctuations correlated over $\xi$ also requires a quasiparticle mean free path $|\mathbf v|/\Gamma_{\rm out}$ longer than $\xi$, because the Born approximation treats the phase that an electron accumulates while crossing a correlated region as small, and with the Born rate its square is of order $\xi\Gamma_{\rm out}/|\mathbf v|$.

\subsection{The magnetic order changes the final states available to scattering}
For a field $E_j$, the drive is $X_0=Fv_j$ at $\phi_0=0$, and its derivative with respect to $\sigma\phi_0$ is $X_1=Fu_j+d_{\mathbf k}F'v_j$ with $F'=dF/dE$, evaluated at $E=\epsilon_0-\mu$. The first term of $X_1$ changes the velocity in the drive, and the second shifts the occupied states in energy. We write the driven distribution as $\delta f_\sigma=-eE_j\tau_{\rm bg}(h_++\sigma\phi_0h_-)$, where $h_+$ is the response common to both spins and $h_-$ the part that differs between them.

Let $\mathcal M_+$ and $\mathcal M_-$ relax the common and opposite parts of the distribution on the unsplit band, including the unit background contribution, and let $\mathcal M_1$ be the derivative of the collision operator with respect to $\phi_0$. The collision operator is invariant under $\phi_0\to-\phi_0$ combined with exchange of the spin labels, so its diagonal blocks are even in $\phi_0$ and its off diagonal blocks are odd. The derivative at $\phi_0=0$ is purely off diagonal and couples the common part to the spin difference. Differentiating the kinetic equation at fixed density, temperature and fluctuation strengths gives
\begin{equation}
h_+=\Mcal_+^{-1}X_0,\qquad
h_-=\Mcal_-^{-1}(X_1-\Mcal_1h_+).
\label{eq:kinetic}
\end{equation}
Two terms generate the spin-dependent response: the change in the driving term, \(X_1\), and the order-dependent collision term, \(-M_1h_+\). The latter converts part of the spin-independent response into a spin imbalance.

Differentiating the elastic constraint at fixed fluctuation strengths gives
\begin{align}
 (\Mcal_1h)_{\kk}&=\tau_{\rm bg}\int_{\kk'}[\mathcal A_L(d_\kk-d_{\kk'})
 +\mathcal A_T(d_\kk+d_{\kk'})]\nonumber\\
 &\qquad\times\delta'(\epsilon_0(\kk)-\epsilon_0(\kk'))(h_\kk-h_{\kk'}),
 \label{eq:directM1}
\end{align}
where $\mathcal A_{L,T}=A_{L,T}|\Lambda_m|^2/(\xi^{-2}+q^2)$ contain the dimensional rate prefactors. The $h_\kk$ and $-h_{\kk'}$ terms describe scattering out of and into the state $\kk$. After integration over the final energy, $\delta'$ produces a positive energy derivative of the integral over the final energy shell, acting on $h$ including its thermal factor.

The splitting displaces the two spin Fermi lines, which makes the accessible final states spin dependent. In a transition $(\mathbf k,\sigma)\to(\mathbf k',\sigma')$, the unsplit energies of the initial and final states differ by $\sigma\phi_0(d_{\mathbf k}-\sigma\sigma'd_{\mathbf k'})$. 
For nearly forward spin-conserving scattering, \(d_k-d_{k'}\simeq 0\), so the contribution to \(M_1\) is small. By contrast, a nearly forward spin flip involves \(d_k+d_{k'}\simeq 2d_k\), giving a finite order-dependent energy shift. $\mathcal M_1$ describes how the magnetic order changes which final states the scattering can reach, and in this way the order turns part of the common response into a spin imbalance. Supplemental Material Sec.~S3 gives the form on energy contours used in the numerics.

For current direction $i$ and field direction $j$ we define the velocity, occupation, charge and total spin coefficients
\begin{align}
V&=\int_\kk u_i h_+,&P&=\int_\kk v_i h_-,\nonumber\\
C&=\int_\kk v_j h_+,&S&=V+P,
\label{eq:responses}
\end{align}
with the order amplitude and background time factored out. In two dimensions each coefficient has units of energy divided by $\hbar^2$. The conductivities are
\begin{equation}
\sigma^s_{ij}=2e^2\tbg\phi_0 S,\qquad
\sigma^c_{jj}=2e^2\tbg C,
\label{eq:conductivities}
\end{equation}
so the spin splitter ratio is $\theta=\phi_0Q$ with the reduced spin splitter ratio $Q=S/C$.

With background scattering alone, $K=a=0$, the velocity, occupation and charge coefficients follow directly from the electric drive,
\begin{align}
C^V_{ij}&=\int_\kk Fu_i v_j,\nonumber\\
C^P_{ij}&=\int_\kk v_i(Fu_j+d_\kk F'v_j)
=-\frac1\hbar\int_\kk d_\kk F\,\partial_{k_j}v_i,\nonumber\\
C_{0,jj}&=\int_\kk Fv_j^2,
\label{eq:meanfield}
\end{align}
where the second form of $C^P$ follows by integration by parts over the periodic zone, or in a continuum with vanishing boundary terms. The reduced spin splitter ratio in this limit is $Q_{\rm CRT}=(C^V+C^P)/C_0$, the value for a constant relaxation time. For the square lattice, symmetry under $k_x\leftrightarrow k_y$ removes the $t_2$ term from the curvature integral and turns the $t_1$ term into a square,
\begin{equation}
 C^P_{xx}=-\frac{t_1d_0}{\hbar^2}\int_\kk F(\cos k_x-\cos k_y)^2.
 \label{eq:latticeidentity}
\end{equation}
The bare occupation coefficient is a density of states weighted by the squared form factor, with the sign of $-t_1d_0$. For $|t_2|<t_1/2$ and positive $d_0$, $C^V$ is also negative, so already with a constant relaxation time the occupation contribution is a sizable term that adds to the velocity contribution. On the NN band this gives $|Q_{\rm CRT}|=0.966$, nearly twice $|b|=1/2$.

\subsection{Spin and charge current decay at weak magnetic scattering}\label{sec:weak}
To identify what controls how the spin splitter ratio changes with disorder, we expand Eq.~\eqref{eq:kinetic} to first order in the magnetic rates, using $\Mcal_\pm^{-1}=I-\tau_{\rm bg}\Lcal_\pm+O(\Lcal^2)$ and $\Mcal_1=\tau_{\rm bg}\Lcal_1$. With $X_0=Fv_x$, $X_1=Fu_x+d_{\mathbf k}F'v_x$ and $S_0=C^V+C^P$, the charge and spin coefficients have the initial decay rates
\begin{align}
\Gamma_c&=\langle v_x,\Lcal_+X_0\rangle/C_0,\nonumber\\
\Gamma_s&=[\langle u_x,\Lcal_+X_0\rangle+\langle v_x,\Lcal_-X_1\rangle+\langle v_x,\Lcal_1X_0\rangle]/S_0.
\label{eq:prb_rates}
\end{align}
$\Gamma_c$ and $\Gamma_s$ are the rates at which the charge current and the spin current start to decay when magnetic scattering is switched on, averaged over the distribution driven by the field. To first order, $C_0/C-1=\tau_{\rm bg}\Gamma_c$ and $1-Q/Q_{\rm CRT}=\tau_{\rm bg}(\Gamma_s-\Gamma_c)$. With $x=C_0/C-1=\rho_{\rm mag}/\rho_{\rm bg}$ the relative magnetic resistivity and $Y=1-Q/Q_{\rm CRT}$ the relative reduction of the reduced spin splitter ratio, $Y=sx$ to first order, with
\begin{equation}
s=\frac{\Gamma_s}{\Gamma_c}-1.
\label{eq:slope}
\end{equation}
The initial slope of the spin splitter ratio against the relative magnetic resistivity thus equals the relative excess of the spin decay rate over the charge decay rate. A spin current that decays faster gives $s>0$, and the ratio falls as magnetic scattering grows. Equal rates give $s=0$, and a faster decay of the charge current gives $s<0$ and a rising ratio. $\Gamma_c$ is positive because $\Lcal_+$ is positive semidefinite, whereas $\Gamma_s$ can have either sign. The $\Lcal_1$ term enters only the spin rate, because the charge conductivity is even in $\phi_0$.

\begin{figure}[tbp]\centering\includegraphics[width=3.4in]{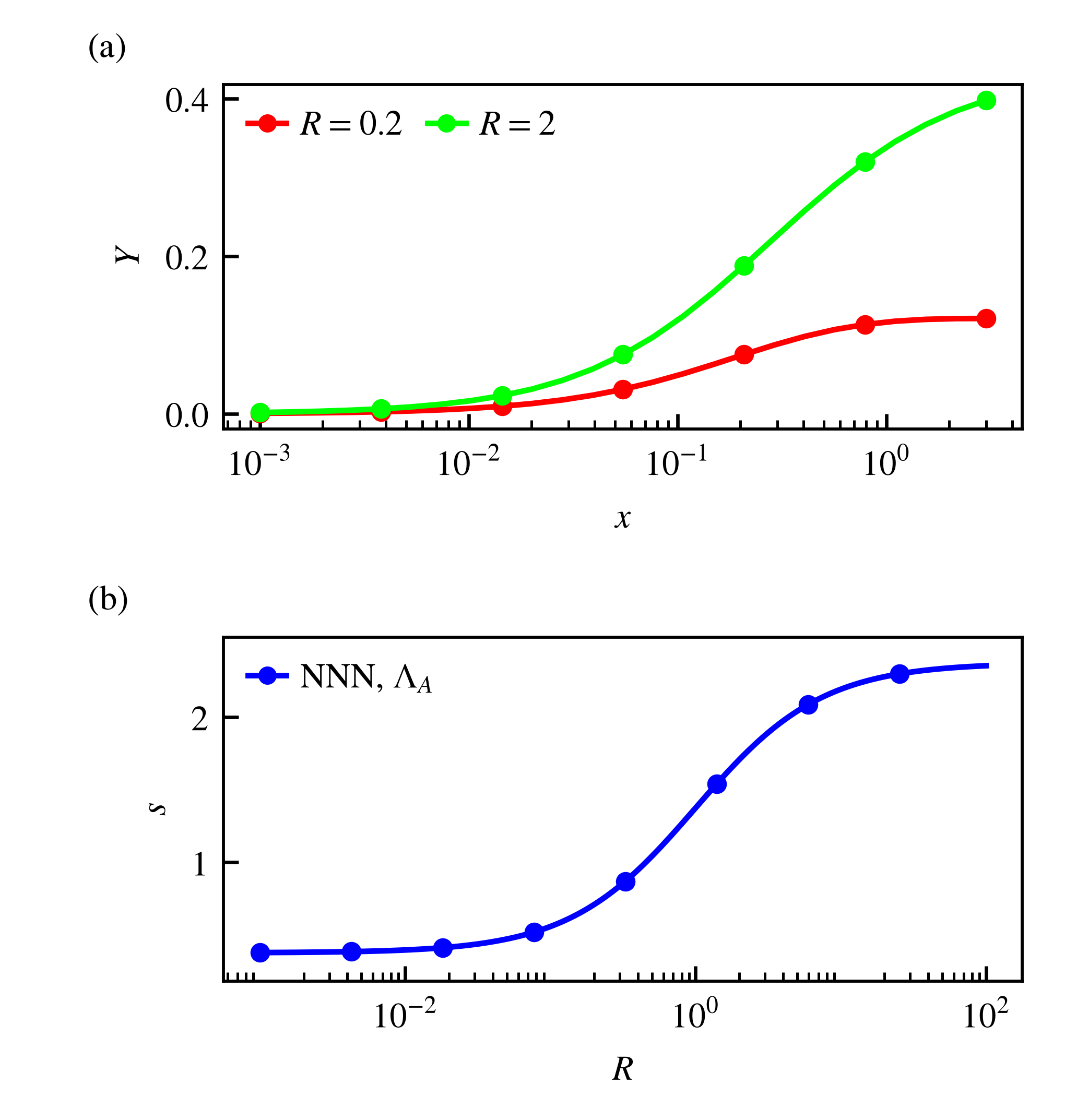}
\caption{Equal charge transport with different spin conversion on the NNN reference band for $\Lambda_A$ at fixed density, $k_BT/t_1=0.05$ and $\xi\bar k_F=4$. (a) Relative reduction $Y=1-Q/Q_{\rm CRT}$ of the reduced spin splitter ratio against relative magnetic resistivity $x$. Red and green denote $R=0.2$ and $2$. At each total strength $A_L+A_T$, the two mixtures of spin conserving and spin flip scattering give the same charge response. (b) Slope $s$ as a function of $R$. Figure~2 of the companion Letter \cite{JiangLetter2026} shows the NN band.}
\label{fig:chargeambiguity}\end{figure}

At zero splitting, both kinds of fluctuations scatter an electron out of a state at the same rate and into it from the same states, apart from their strengths. We call $D$ the part of the collision operator describing scattering out of a state and $G$ the part describing scattering into it. A spin flip exchanges the spin labels, so transverse scattering into a state enters the spin difference with the opposite sign,
\begin{align}
\Lcal_+&=(A_L+A_T)(D-G),\nonumber\\
\Lcal_-&=A_L(D-G)+A_T(D+G),\nonumber\\
s(R)&=\frac{s_L+Rs_T}{1+R}.
\label{eq:prb_channels}
\end{align}
The last line follows because $\Gamma_c$ and $\Gamma_s$ are linear in the strengths. Here $s_L$ is the slope for purely spin conserving scattering and $s_T$ the slope for purely spin flip scattering, and $s$ moves continuously from $s_L$ to $s_T$ as the spin flip fraction $R/(1+R)$ grows. At fixed $A_L+A_T$, $\Lcal_+$ stays the same, so the charge conductivity and the spin averaged scattering rates are independent of $R$, while $\Lcal_-$ and $\Lcal_1$ change at first order in the splitting. Figure~2 of the companion Letter \cite{JiangLetter2026} shows this for the NN band and Fig.~\ref{fig:chargeambiguity} for the NNN band. In Fig.~\ref{fig:chargeambiguity}(a) the curves share one resistivity axis and give different spin conversion, and Fig.~\ref{fig:chargeambiguity}(b) shows $s(R)$. For both couplings on the two lattice reference bands, $s_T=2.37$ to $7.53$ and $|s_L|\le0.74$. On the NN band $s_L<0<s_T$, so $s$ changes sign at $R^*=-s_L/s_T$, which is $0.133$ for $\Lambda_A$. On the NNN band both limits are positive. Across a wide sample of band parameters $s_T$ is always positive, so a negative slope requires $s_L<0$ and $R<R^*$. Negative slopes occur in 8 of 864 grid sets and 62 of 1334 random sets (Supplemental Material Sec.~S4).

\begin{figure}[tbp]\centering\includegraphics[width=3.4in]{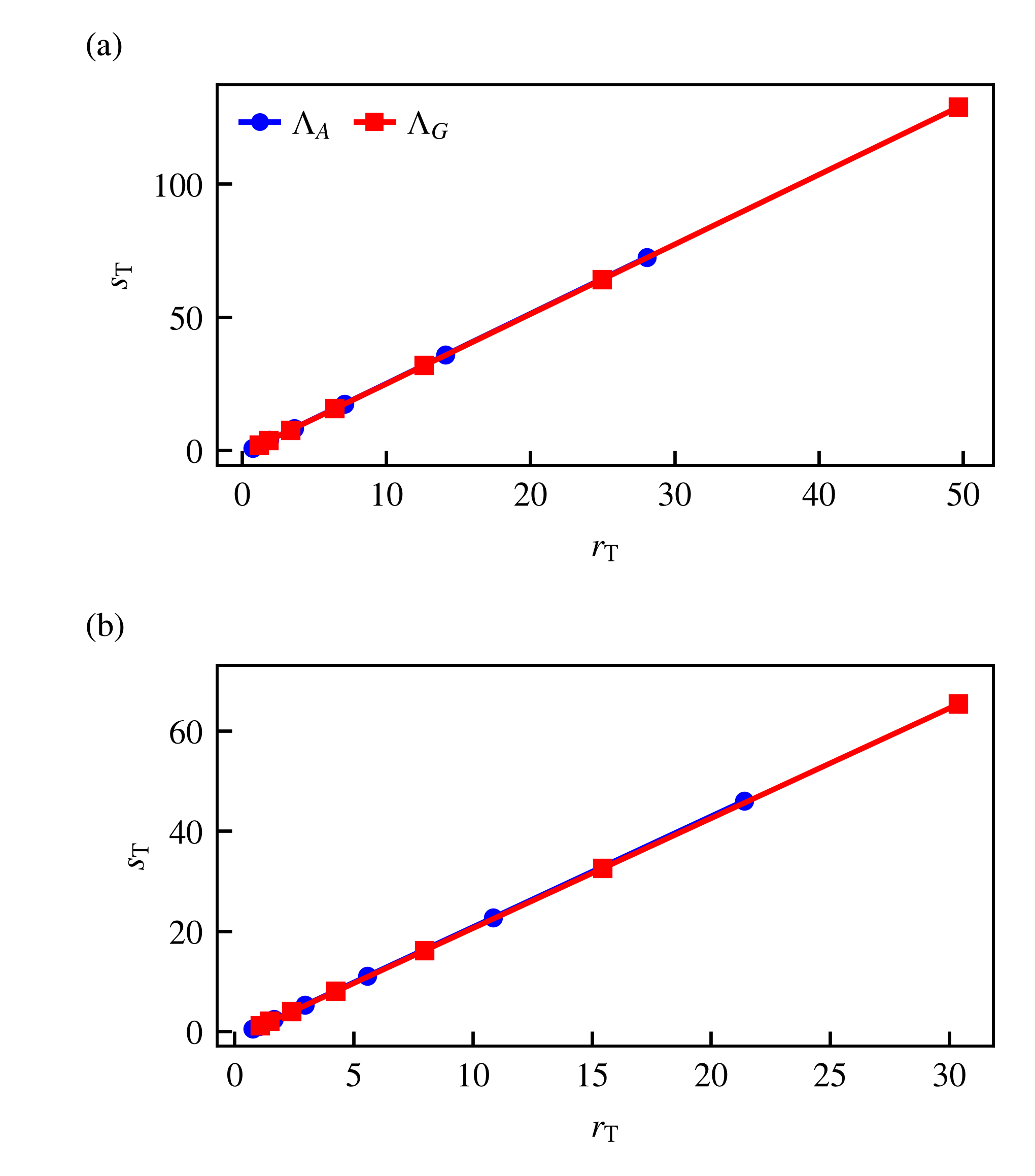}
\caption{Pure spin flip slope $s_T$ against $r_T=\Gamma^{\rm avg}_{T,\rm out}/\Gamma_{T,\rm tr}$, with $\Gamma_{T,\rm tr}$ defined in Sec.~\ref{sec:weak}, on (a) the NN and (b) the NNN reference band, at fixed density and $k_BT/t_1=0.05$. The outgoing rate is averaged over the unsplit Fermi line with density of states weights. Blue circles denote $\Lambda_A$ and red squares $\Lambda_G$, for $\xi\bar k_F=1,2,4,8,16,32,64$. The band integral of Appendix~\ref{sec:transverse} gives $\alpha=2.65$ on NN and $2.32$ on NNN.}\label{fig:forwardrates}\end{figure}

A nearly forward spin flip reverses the spin and keeps the direction of motion almost unchanged, so it damages the charge current little. The field drives each spin with its own velocity, so at equal energy the two spin populations differ in proportion to the velocity difference of the two spins there. The flip conserves energy, so it moves the electron to the other spin band at the same energy and strongly relaxes this imbalance. For the same reason, the change of the collision process with the order cancels the spin flip relaxation of the part of the spin difference that only shifts the band energy [Eq.~\eqref{eq:thermalcancel}], and this protection holds at every scattering strength [Eq.~\eqref{eq:exactsolution}]. The spin flip slope grows with $r_T=\Gamma^{\rm avg}_{T,\rm out}/\Gamma_{T,\rm tr}$, the ratio of the spin flip quasiparticle scattering rate to $\Gamma_{T,\rm tr}$, twice the charge decay rate of purely spin flip scattering on the unsplit Fermi line, the normalization used for $K$, as $s_T\simeq\alpha r_T+O(1)$, with $\alpha$ fixed by the band [Eq.~\eqref{eq:forwardalpha}]. The band integrals give $\alpha=2.65$ for the NN band and $2.32$ for the NNN band (Fig.~\ref{fig:forwardrates}). In two dimensions $r_T$ grows linearly with $\xi$, and on a three dimensional Fermi surface only logarithmically (Appendix~\ref{sec:transverse}). Supplemental Material Sec.~S5 treats unequal longitudinal and transverse correlation lengths.

\section{Microscopic origin of the slope}\label{sec:microscopic}
\subsection{Lifetime and transport time approximations}\label{sec:totalratio}
All models below keep the band, density, temperature, couplings and scattering strengths fixed. The reference $Q_{\rm CRT}=(C^V+C^P)/C_0$ contains the background alone. The spin dependent quasiparticle lifetime, the analog of the self energy relaxation time approximation (SERTA) \cite{Ponce2020}, is $\tau_\sigma=\tau_{\rm bg}/(1+\tau_{\rm bg}\Gamma_{{\rm out},\sigma})$. The transport time approximation, the analog of the momentum relaxation time approximation (MRTA), uses
\begin{align}
 \gamma_{{\rm tr},\sigma}(\kk)&=\sum_{\kk'\sigma'}W_{\kk\sigma,\kk'\sigma'}
 \left[1-\frac{\mathbf v_{\kk'\sigma'}\cdot\mathbf v_{\kk\sigma}}{|\mathbf v_{\kk\sigma}|^2}\right],\nonumber\\
 \tau_{{\rm tr},\sigma}(\kk)&=\frac{\tau_{\rm bg}}{1+\tau_{\rm bg}\gamma_{{\rm tr},\sigma}(\kk)}.
\label{eq:transporttime}
\end{align}
Both approximations use the split bands and replace scattering into a state, which connects different momenta, by a single local rate. They lose the angular and spin structure of the relaxation.

A further model keeps the full $\Mcal_\pm$ and sets $\Lcal_1=0$. Its collisions conserve the unsplit energy while the drift keeps the split bands, so it isolates the effect of energy conservation on the split bands. We call it the unsplit collision model.

Over a range of scattering strengths on the NN, NNN and parabolic bands, the full calculation reduces $|Q|$ by 3 to 62 percent and keeps its sign. The change of the collision process with the order carries a large part of this reduction. Where it carries more than the whole reduction, the unsplit collision model and the full calculation change $Q$ in opposite directions. This happens at weak scattering on the NN band and throughout the parabolic band, and nowhere on the NNN band. On the lattice, the change of scattering out of a state recovers most of the correction, whereas in the parabolic band the change of scattering into a state is essential to the suppression.

At weak scattering, the transport time approximation reproduces the initial growth of the resistivity exactly. To first order in the rates, the full charge coefficient changes by $-\tau_{\rm bg}\sum_{\kk\kk'}FW(v_x^2-v_xv_x')$ and the transport time coefficient by $-\tau_{\rm bg}\sum_{\kk\kk'}FWv_x^2(1-\mathbf v\cdot\mathbf v'/|\mathbf v|^2)$, where $W$ is the transition rate. Because $F_{\mathbf k}=F_{\mathbf k'}$ for elastic transitions, $FW$ is symmetric, and because the unsplit scattering is invariant under $k_x\leftrightarrow k_y$,
\begin{equation}
\sum FWv_x^2\frac{\mathbf v\cdot\mathbf v'}{|\mathbf v|^2}=\frac12\sum FW\,\mathbf v\cdot\mathbf v'=\sum FWv_xv_x',
\label{eq:mrta_charge}
\end{equation}
so the two charge slopes coincide. The average of $\gamma_{\rm tr}$ weighted by $Fv_x^2$ is exactly the charge decay rate $\Gamma_c$, whereas $\Gamma_{T,\rm tr}$ in $r_T$ and the strength $K$ are defined as twice the corresponding charge decay rate on the Fermi line.

The transport time approximation gives 23 to 47 percent of the slope $s$ on the reference bands, whereas the lifetime approximation overestimates the growth of the resistivity by factors of $2.2$ to $5.4$. At strong scattering, the transport time approximation overestimates the correction to $Q$ by up to a factor of $4.3$ and predicts a sign change of $Q$ at nine parameter points where the full calculation keeps the sign (Supplemental Material Sec.~S6).

\subsection{Spin flips drive \texorpdfstring{$Q$}{Q} to the band ratio on the NN band}\label{sec:decomposition}
\begin{figure}[tbp]
\centering
\includegraphics[width=3.4in]{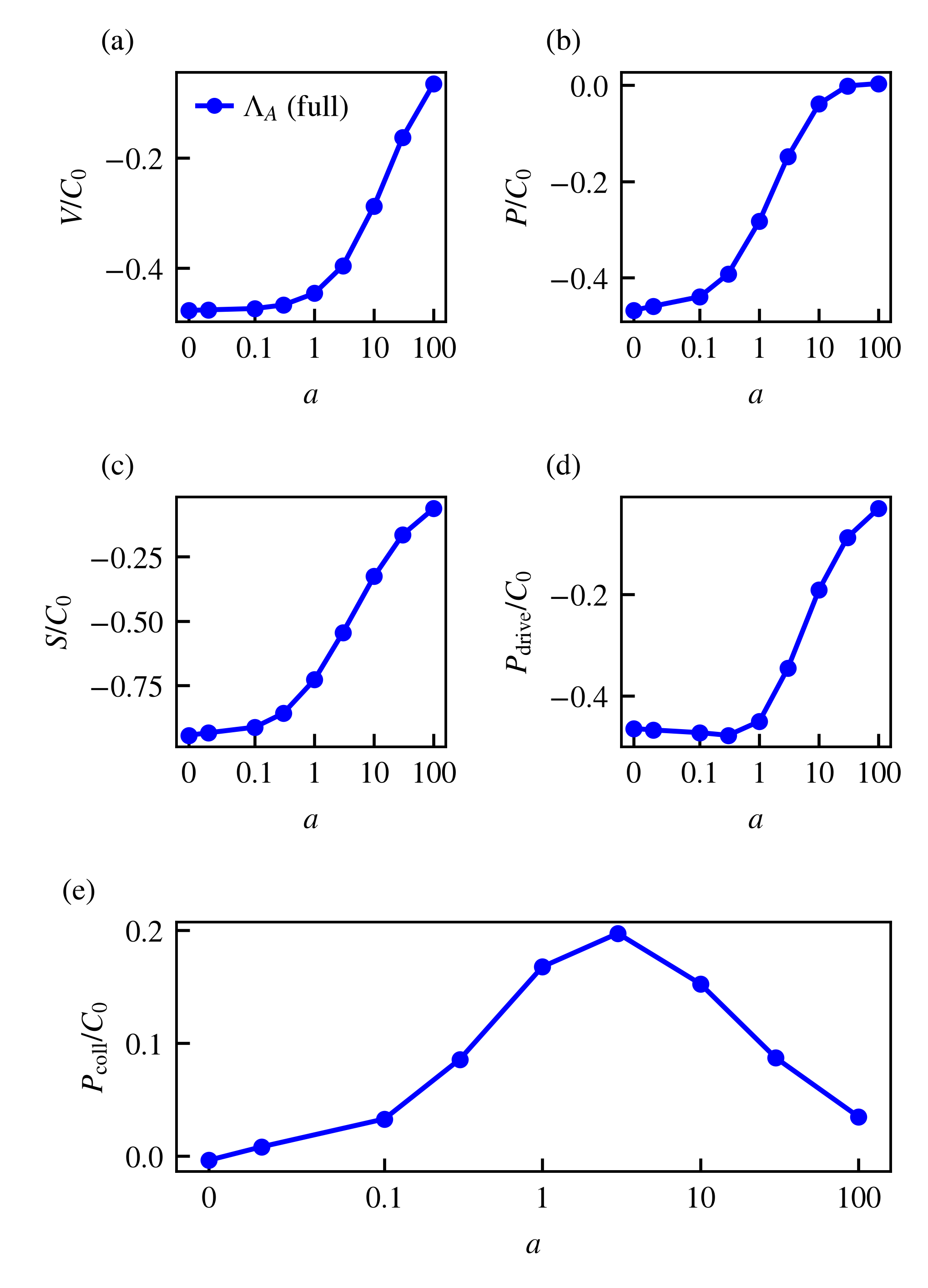}
\caption{Velocity and occupation contributions in the momentum partition on the NN reference band with $\Lambda_A$, at fixed density, $K=0.1$, $k_BT/t_1=0.05$ and $\xi\bar k_F=4$. Panels (a) to (e) show $V/C_0$, $P/C_0$, $S/C_0$, $P_{\rm drive}/C_0$ and $P_{\rm coll}/C_0$, respectively, as functions of spin flip strength $a$. The normalization $C_0$ is the charge coefficient at $K=a=0$. The horizontal axis is linear for $a\le0.1$ and logarithmic above that value.}
\label{fig:microscopic}
\end{figure}

On the NN band $u_x=-v_x/2$, so $V=bC$ at every scattering strength [Fig.~\ref{fig:microscopic}(a)] and $Q=-1/2+P/C$. As $a$ grows at fixed $K=0.1$, $P/C_0$ rises from $-0.47$ to nearly zero [Fig.~\ref{fig:microscopic}(b)], so spin flips drive $Q$ from its constant relaxation time value $-0.97$ to about the velocity only value $b=-1/2$. The velocity contribution remains negative and larger, so the total $S$ keeps its sign [Fig.~\ref{fig:microscopic}(c)]. For $\Lambda_A$ the occupation term vanishes near $a\simeq33$, where $R\simeq49$ and $x\simeq2.2$, and $P$ then changes sign and stays below $0.01C_0$ for $0\le a\le100$. For $\Lambda_G$ the sign change occurs near $a\simeq11$ (Supplemental Material Sec.~S7). A measured $Q$ above $-1/2$ signals a positive occupation contribution once $\phi_0$ is fixed from the band splitting.

The approach to $b$ comes from the change of the collision process with the order. We split $P$ into the part driven directly by the field, $P_{\rm drive}$, the response to $X_1$, and the part produced by the collisions, $P_{\rm coll}=P-P_{\rm drive}$, the response to $-\Mcal_1h_+$. The field driven part keeps the sign set by the band for $0\le a\le100$ [Fig.~\ref{fig:microscopic}(d)], and the collision part grows to exceed it in magnitude at large $a$ [Fig.~\ref{fig:microscopic}(e)]. The cancellation survives with spin conserving scattering switched off, so spin flip scattering alone drives it (Supplemental Material Sec.~S7). The vanishing of $P$ is a statement about the momentum partition, because the energy partition divides the same total differently (Supplemental Material Sec.~S8), whereas the approach of $Q$ to $b$ is observable. In the parabolic band the occupation contribution stays negative through $a=30$.

\subsection{Dependence of the spin flip limit on the correlation length}\label{sec:xi_dependence}
\begin{figure}[tbp]
\centering\includegraphics[width=3.4in]{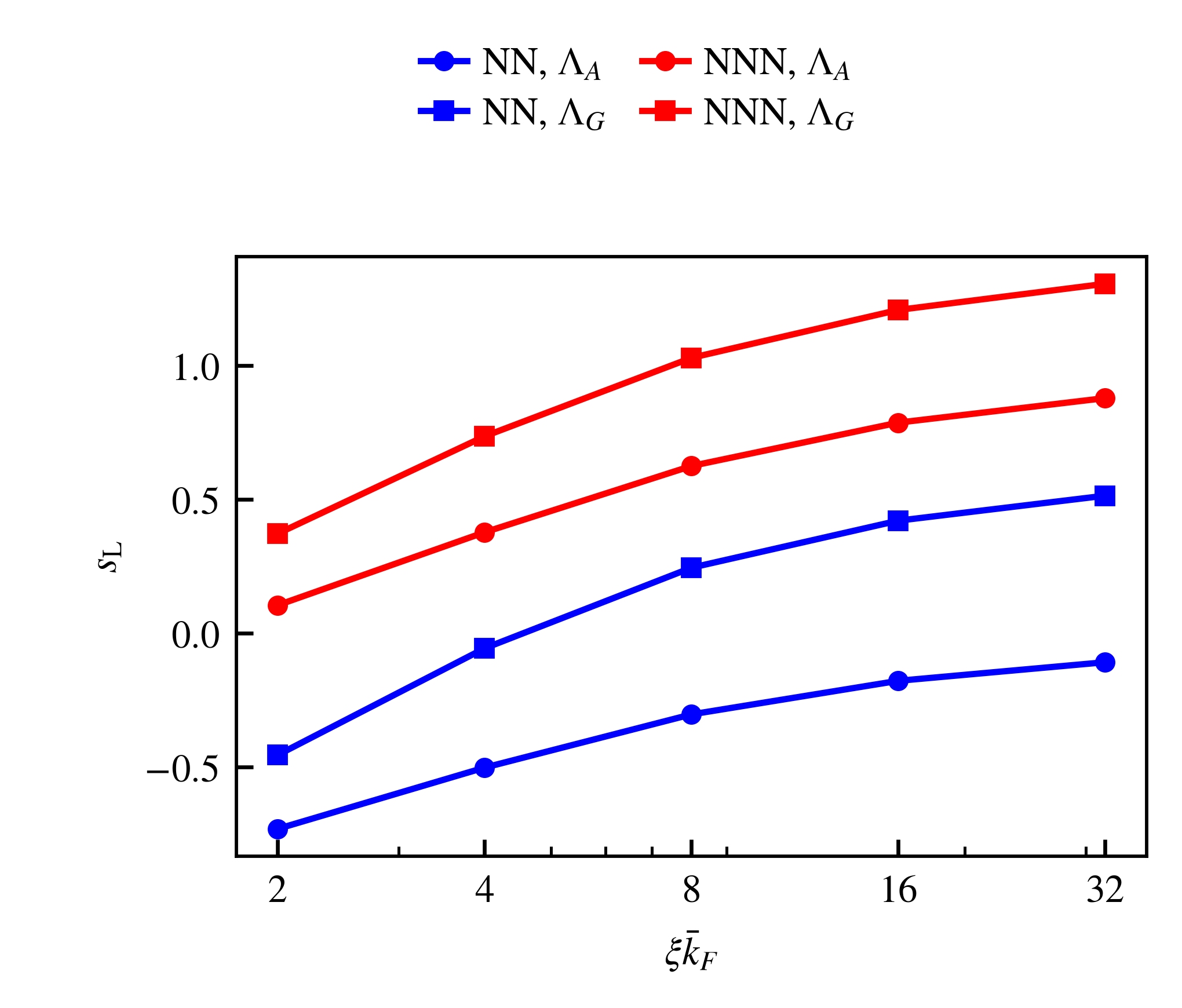}
\caption{Spin conserving slope $s_L$ against correlation length on the NN and NNN reference bands at fixed density and $k_BT/t_1=0.05$. Blue denotes NN and red NNN. Circles denote $\Lambda_A$ and squares $\Lambda_G$. Figure~1 of the companion Letter \cite{JiangLetter2026} shows $r_T$ and $s_T$ for the same bands and couplings.}\label{fig:xi_dependence}
\end{figure}

Figure~1 of the companion Letter \cite{JiangLetter2026} follows $r_T$ and $s_T$ on the two reference bands and both couplings over $\xi\bar k_F=2$ to $32$, with the rates recalibrated at each correlation length and the density fixed, and Fig.~\ref{fig:xi_dependence} shows $s_L$ for the same cases. For Ornstein--Zernike fluctuations, the integral for scattering out of a state grows with $\xi$, while the transport weight suppresses the small momentum transfers responsible for this growth. The transport rate approaches a finite value, so in two dimensions $r_T$ and $s_T$ grow nearly linearly with $\xi$ (Appendix~\ref{sec:transverse}), while on a three dimensional Fermi surface the growth is logarithmic. On the lattice bands the spin conserving slope stays bounded, $|s_L|\le1.31$, because for spin conserving scattering $\Lcal_-=\Lcal_+$ and $\Lcal_1$ carries the weight $d_{\mathbf k}-d_{\mathbf k'}$, which vanishes for forward scattering, so $\Gamma_s$ and $\Gamma_c$ are both transport weighted rates.

On NN, the negative longitudinal slope of $\Lambda_A$ shrinks in magnitude, moving the threshold $R^*=-s_L/s_T$ from $0.444$ at $\xi\bar k_F=2$ to $0.0030$ at $32$. For $\Lambda_G$ the longitudinal slope becomes positive between $4$ and $8$, so $s$ is positive at every $R$ for $\xi\bar k_F\ge8$. Negative slopes are restricted to baths dominated by longitudinal fluctuations. Both NNN channel slopes remain positive over this range.

\begin{figure}[tbp]\centering\includegraphics[width=3.4in]{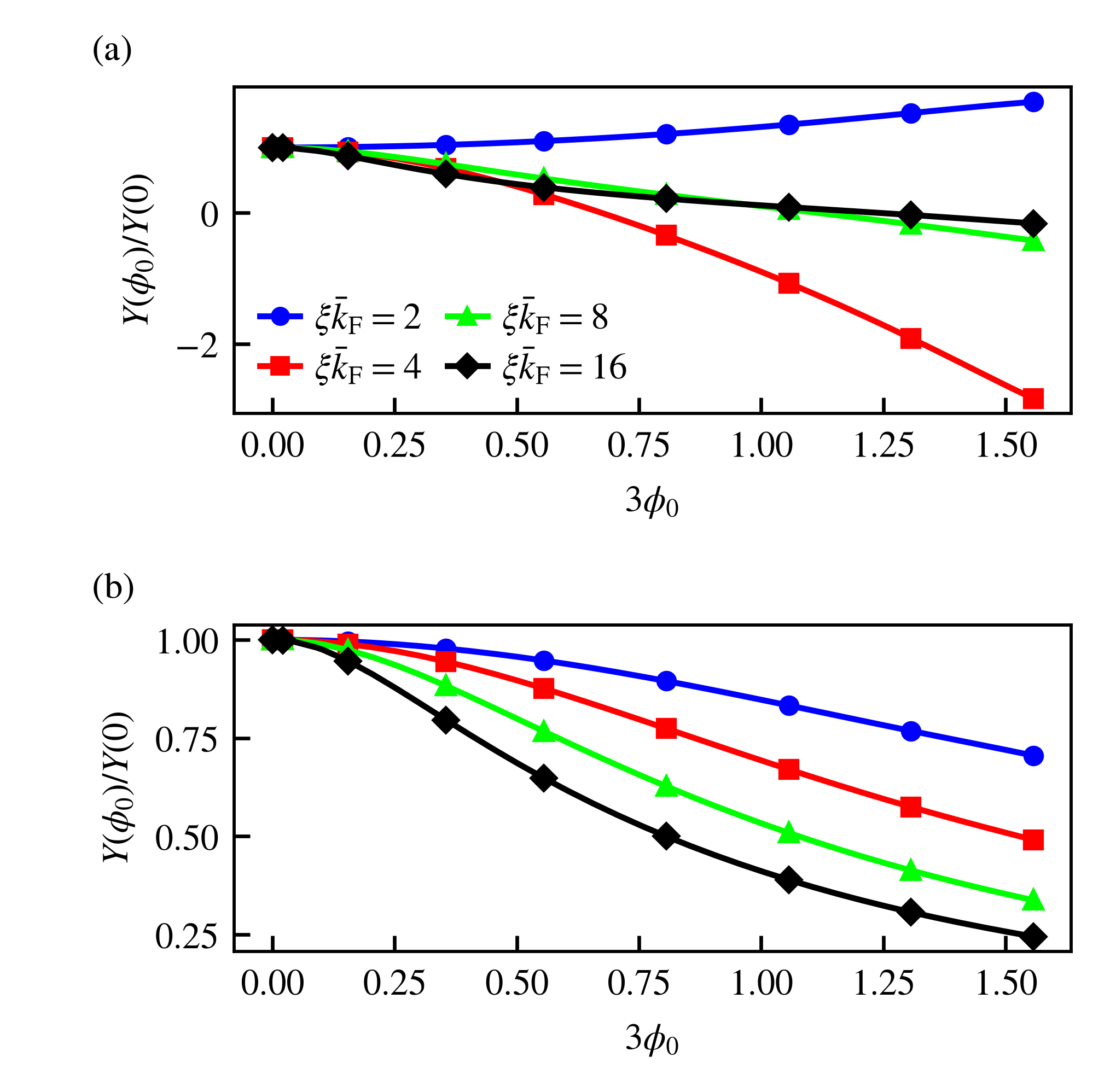}
\caption{Relative reduction of the reduced spin splitter ratio at finite splitting, $Y(\phi_0)=1-\theta(\phi_0)/\theta_{\rm bg}(\phi_0)$, normalized by its limit at $\phi_0\to0$, against $3\phi_0$ on the NN reference band with $\Lambda_A$ and $k_BT/t_1=0.05$. Panels (a) and (b) use $R=0.2$ and purely transverse fluctuations, respectively, at the same total strength, calibrated as $K(1+R)=0.09$, equal to that of $K=0.03$ and $R=2$ at each $\xi$. Blue circles, red squares, green triangles and black diamonds denote $\xi\bar k_F=2,4,8,16$. In (a), $Y(0)$ is negative at $\xi\bar k_F=2$ and positive at $4,8,16$. The chemical potential is solved at fixed density, and at the common endpoint the first spin saddle lies $3k_BT$ from it. Figure~3(a) of the companion Letter \cite{JiangLetter2026} shows $R=2$.}\label{fig:windownonlinear}\end{figure}

A flip between the split Fermi lines requires a momentum transfer of $2\phi_0|d_{\mathbf k}|/(\hbar|\mathbf v|)$. Once this exceeds $1/\xi$, the splitting itself blocks the forward flips. Equivalently, while an electron crosses a correlated region, its spin precesses by more than a radian in the exchange field, so it follows the slowly varying fluctuation axis adiabatically. The blocking is lifted only near the nodes, where both the splitting and the $d$-wave coupling vanish. It also requires that fluctuations at this momentum transfer carry an energy small compared with the splitting, which for magnons means a magnon velocity well below $|\mathbf v|$. At $R=0.2$, a larger splitting turns the suppression into an enhancement [Fig.~\ref{fig:windownonlinear}(a)]. For $R=2$, shown in Fig.~3(a) of the companion Letter \cite{JiangLetter2026}, and for purely transverse fluctuations [Fig.~\ref{fig:windownonlinear}(b)], the reduction of $\theta$ decreases monotonically with splitting, more strongly at longer correlation length. At $3\phi_0t_1=1.5t_1$ and $R=2$, the reduction is $0.53$, $0.37$, $0.26$ and $0.18$ of its value at zero splitting for $\xi\bar k_F=2$, $4$, $8$ and $16$. For purely transverse fluctuations the corresponding values are $0.72$, $0.51$, $0.35$ and $0.26$.

On the NN band the largest splitting on the Fermi line is $\Delta_s=3\phi_0t_1$ and $E_F=3t_1$, so the linear window $\phi_0\xi\bar k_F\lesssim0.5$ of Sec.~\ref{sec:survey} reads $\xi\bar k_F\Delta_s/E_F\lesssim0.5$. In three dimensional Ising, XY and Heisenberg criticality the correlation length exponent $\nu$ exceeds the order amplitude exponent $\beta$ \cite{PelissettoVicari2002}, so $\xi\Delta_s\propto(1-T/T_N)^{\beta-\nu}$ grows towards $T_N$ from below, but slowly, with $\beta-\nu\simeq-0.3$. Close to the transition the splitting blocks the forward flips, while further below they dominate in a window that widens as $\Delta_s/E_F$ decreases. As a rough estimate, with Heisenberg exponents, $\Delta_s=\Delta_s(0)(1-T/T_N)^\beta$, $\xi=\xi_0(1-T/T_N)^{-\nu}$ and $\xi_0=1/\bar k_F$, the condition $\xi\bar k_F\Delta_s/E_F\lesssim0.5$ holds up to $\xi\bar k_F\approx30$ for a low temperature splitting $\Delta_s(0)=0.1E_F$, but only up to $\xi\bar k_F\approx3$ for $\Delta_s(0)=0.3E_F$. Altermagnets whose splitting is small compared with the Fermi energy are the natural place to see the growth of $s_T$ with $\xi$.

\subsection{In the parabolic band only scattering moves the ratio away from \texorpdfstring{$b$}{b}}\label{sec:exactcancel}
The parabolic band of Sec.~\ref{sec:model} has $u_x=bv_x$ with $b=d_0/\mu_0$, and each spin band is an anisotropic parabola, $\varepsilon_{\kk\sigma}=\hbar^2[(1+\sigma\phi_0b)k_x^2+(1-\sigma\phi_0b)k_y^2]/(2m)-\mu$, with $1/m_{x\sigma}=(1+\sigma\phi_0b)/m$ and the same density for both spins. A constant relaxation time gives the Drude conductivities $n_\sigma e^2\tau/m_{x\sigma}$, so $\theta=\phi_0b$ exactly, at every $\phi_0$, temperature and density, and every deviation of $Q$ from $b$ comes from scattering. In the momentum partition $V=bC_0$, so the bare occupation coefficient vanishes. In the first form of Eq.~\eqref{eq:meanfield}, $C^P_{xx}$ is the sum of a velocity piece $\int_\kk Fu_xv_x=bC_0$ and an energy shift piece $\int_\kk d_\kk F'v_x^2=-bC_0$, two large terms of opposite sign, with $C_0=n/(2m)$ the two dimensional Drude weight.

Relaxation that depends on angle or energy weights the two cancelling pieces differently, so it lifts the cancellation already with $\Mcal_1=0$. For example, a lifetime $\tau(\varphi)=\tau_{\rm bg}(1+\alpha\cos4\varphi)$ with $|\alpha|<1$ and $\Lcal_1=0$ gives $P/C=-b\alpha/2$. The quasiparticle scattering rate inherits the $\cos4\varphi$ harmonic of $\Phi_d^2$, which gives a negative $\alpha$, so the lifetime approximation enhances $Q$ (Supplemental Material Sec.~S6).

\begin{figure}[tbp]
\centering\includegraphics[width=3.4in]{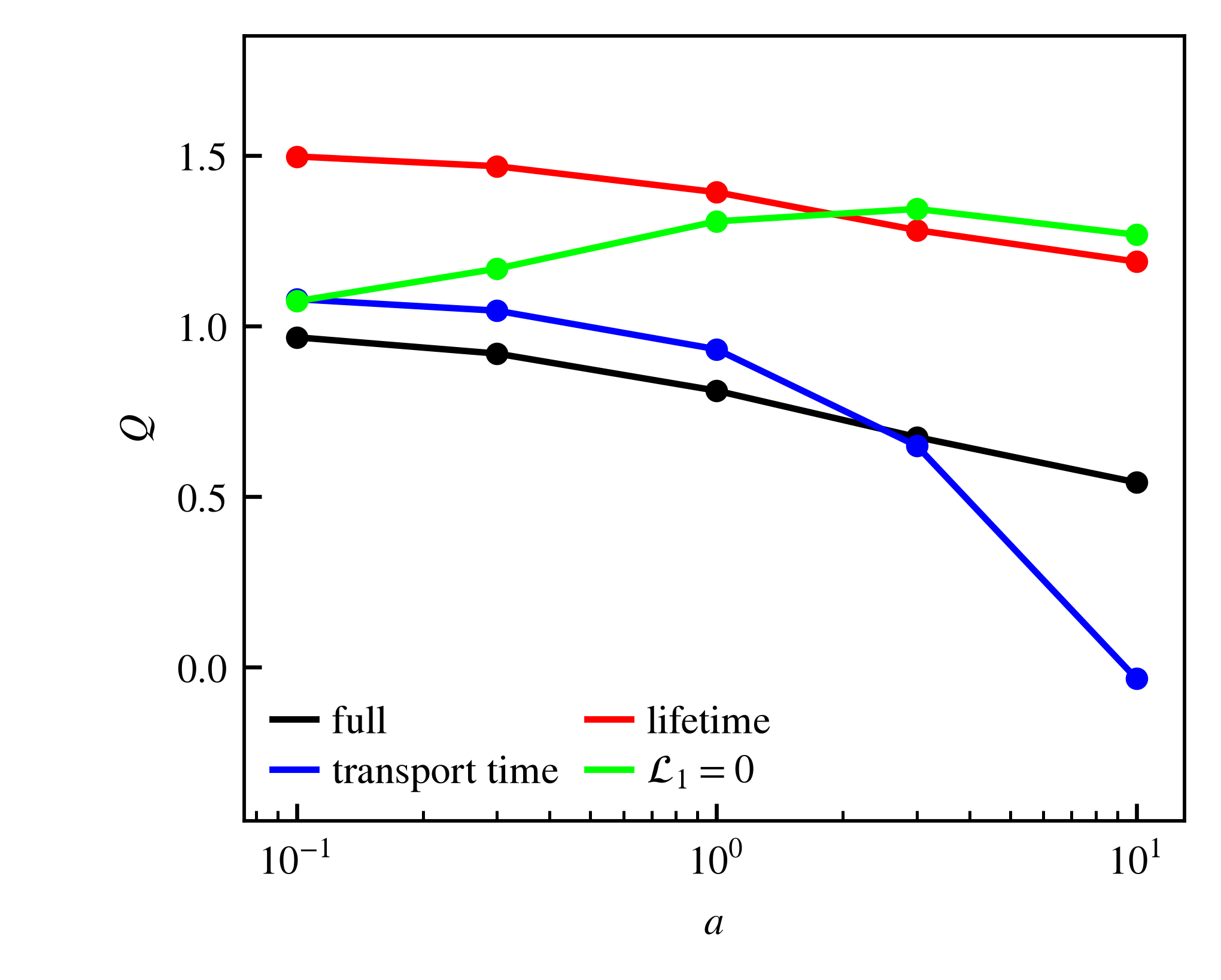}
\caption{Reduced spin splitter ratio $Q=S/C$ against spin flip strength $a$ in the parabolic band for $\Lambda_A$, at fixed density, $b=1$, $K=0.1$, $k_BT/\mu_0=0.05$ and $\xi k_F=4$. Black denotes the full collision operator, blue the transport time approximation, red the lifetime approximation and green unsplit collisions with $\mathcal L_1=0$. The constant relaxation time value is $Q_{\rm CRT}=b=1$.}
\label{fig:parabolic_models}
\end{figure}

In the full calculation, the change of the collision process with the order dominates. At $K=0.1$ and $a=3$, the full calculation gives $Q=0.675$ for $\Lambda_A$, below $Q_{\rm CRT}=b=1$, whereas collisions computed on the unsplit bands give $Q=1.34$ [Fig.~\ref{fig:parabolic_models}]. Collisions on the unsplit bands predict an enhancement, and energy conservation on the split bands turns it into a suppression. The full calculation suppresses $Q$ and the unsplit collision model enhances it also for $\Lambda_G$ and for all six coupling forms at $a=3$ (Supplemental Material Sec.~S6). The change of the collision process with the order and the change of the electric drive enter at the same order in $\phi_0$, so both are part of the linear response.

\subsection{Range of validity}\label{sec:survey}
We repeat the calculation on 72 combinations of hopping ratio, filling $\nu$ (electrons per spin per site) and temperature (Supplemental Material Sec.~S9). Away from the van Hove saddle points, the occupation contribution remains a leading term, about one half of the charge coefficient before scattering and one third at $a=1$.

\begin{figure}[tbp]
\centering\includegraphics[width=3.4in]{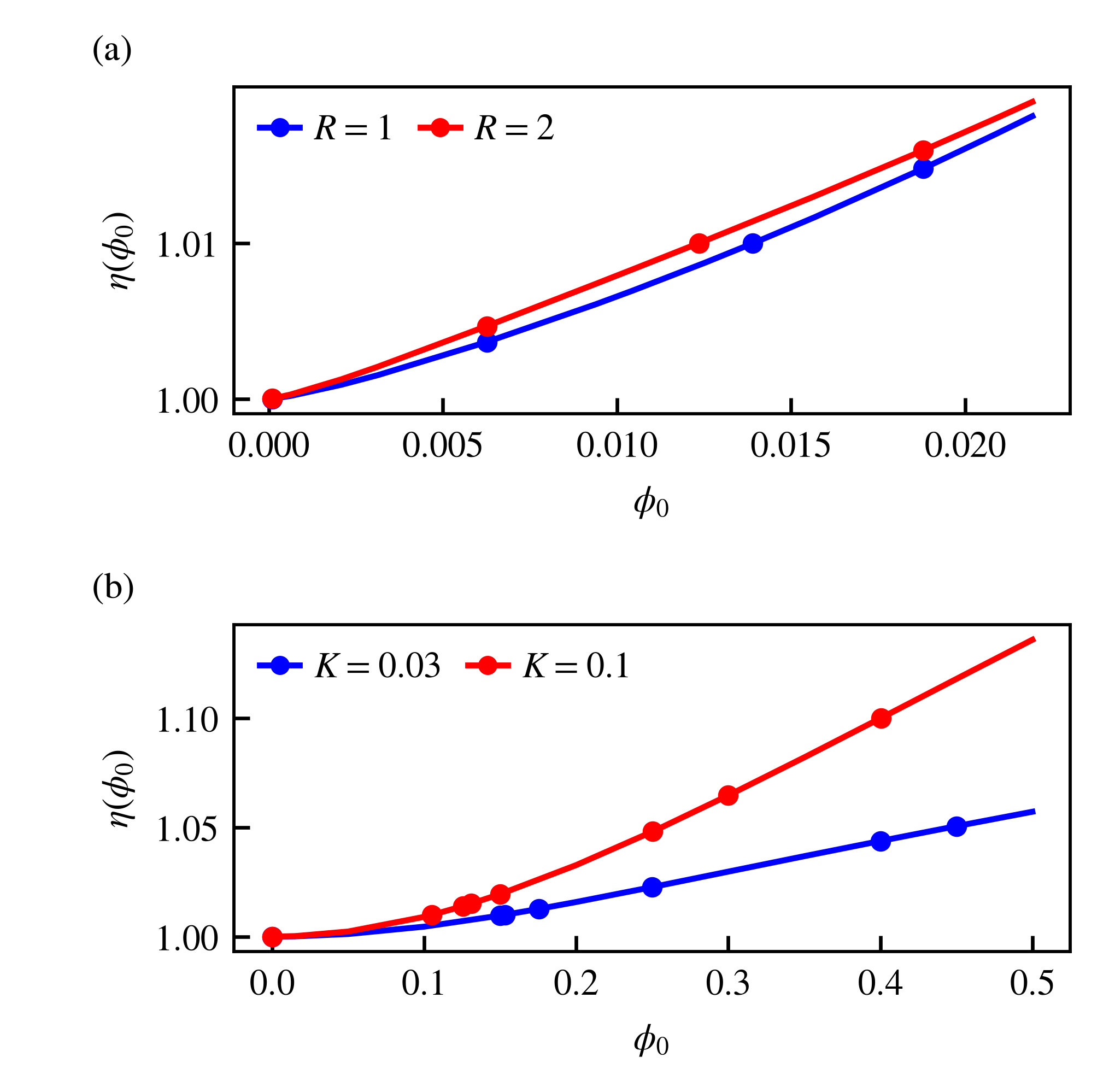}
\caption{Response at finite splitting relative to the prediction at linear order, $\eta(\phi_0)=\theta(\phi_0)/[\theta_{\rm bg}(\phi_0)(Q/Q_{\rm CRT})_{\rm lin}]$, at fixed density with $\Lambda_A$ and $\xi\bar k_F=4$. The background response is evaluated at the same density and splitting. (a) The band with $t_2/t_1=-0.2$, $\nu=0.40$, $k_BT/t_1=0.02$ and $K=0.1$. Blue and red denote $R=1$ and $R=2$. (b) The NN reference band at $k_BT/t_1=0.05$ and $R=2$, with blue for $K=0.03$ and red for $K=0.1$. The full saddle splitting in (a) is $4\phi_0t_1$, and the full splitting on the unsplit Fermi line in (b) is $3\phi_0t_1$.}
\label{fig:article_validity}
\end{figure}

Near the van Hove saddle points, the occupation contribution becomes larger and the linear range in the order amplitude becomes narrower. At the saddles $(\pi,0)$ and $(0,\pi)$, at energy $\epsilon_0=4t_2$, the velocity vanishes and $|d_{\mathbf k}|=2t_1$ is largest, so by Eq.~\eqref{eq:latticeidentity} the saddle region adds strongly to $C^P$ and little to $C_0$. The linear coefficient applies while each spin split saddle, shifted by $2\phi_0d_0$, stays on one side of the chemical potential, $2\phi_0d_0<|\mu(\phi_0)-4t_2|$. Close to the saddle, the measured ratio departs from its linear value by 1 percent at a full splitting of only about $0.05t_1$ [Fig.~\ref{fig:article_validity}(a)], whereas on the NN band the linear range extends to full splittings of $0.32t_1$ to $0.46t_1$ [Fig.~\ref{fig:article_validity}(b)].

The linear range also shrinks as the correlation length grows. A spin flip between the split bands requires a momentum transfer of at least $2\phi_0|d_{\mathbf k}|/(\hbar|\mathbf v|)$, and the linear expansion requires this transfer times $\xi$ to be small. For NN with $\Lambda_A$, $R=2$ and $K(1+R)=0.09$, the window at long correlation length is approximately $\phi_0\xi\bar k_F\lesssim0.5$ (Supplemental Material Sec.~S10). Beyond it the splitting blocks forward flips and reduces the suppression below its linear value (Fig.~\ref{fig:windownonlinear}). Defects with a finite range change the slope by 8 to 15 percent for a range of about $1/\bar k_F$ and much more for longer ranges (Supplemental Material Sec.~S11).

\section{Strong spin flip scattering}\label{sec:strong}
\begin{figure*}[tbp]
\centering\includegraphics[width=\textwidth]{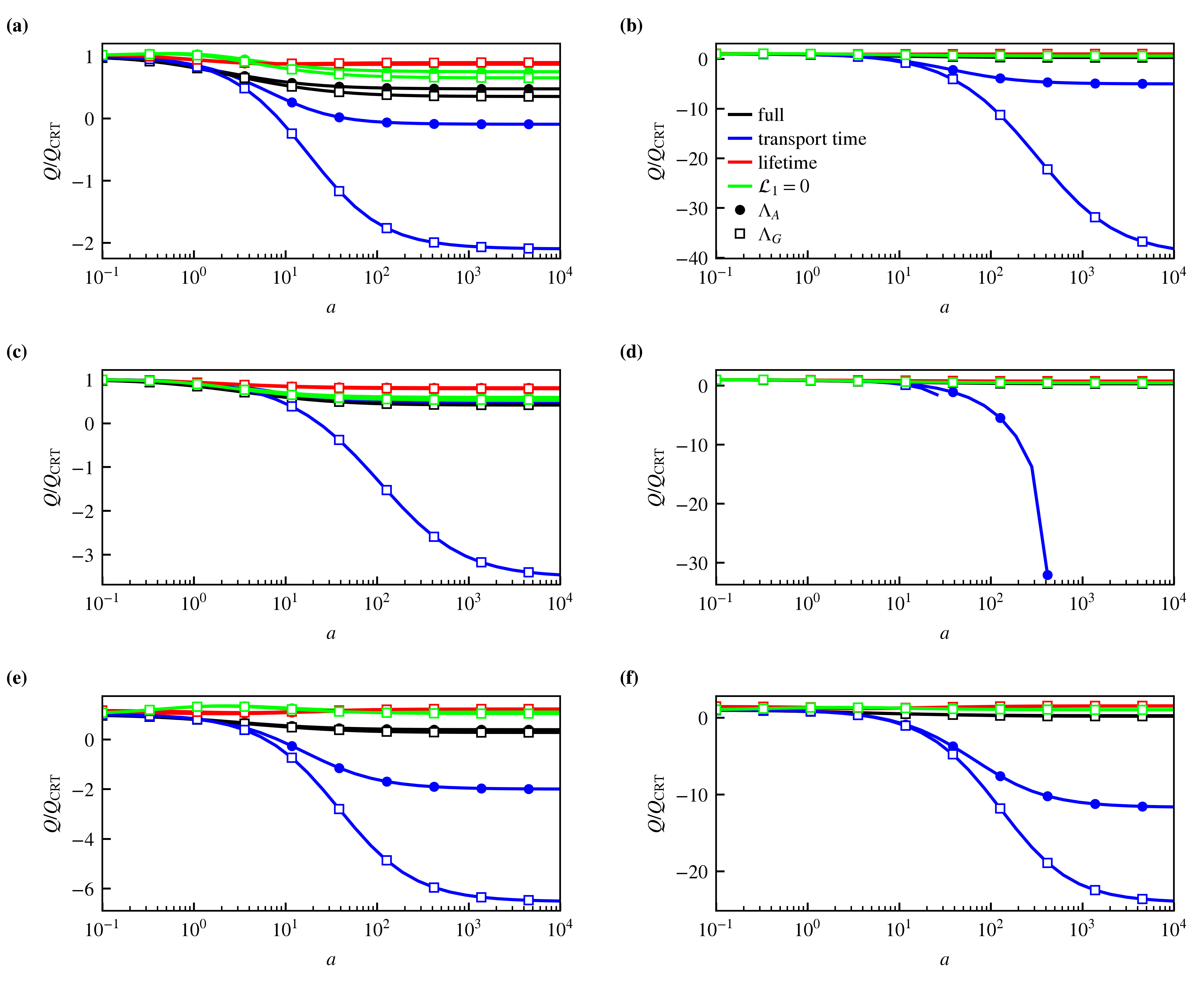}
\caption{Ratio at linear order under strong spin flip scattering at $K=0.01$ and fixed density. The rows use NN, NNN and the parabolic band with $b=1$, and the columns use correlation lengths $4$ and $16$ in units of the inverse mean Fermi radius. Black denotes full collisions, blue transport time, red lifetime and green collisions computed on unsplit bands with $\mathcal L_1=0$. Circles denote $\Lambda_A$ and squares $\Lambda_G$. The vertical normalization is the bare $Q_{\rm CRT}$ of each band. The temperature is $0.05$ in the band energy units. Transport time curves on NNN at $16$ end where their total local relaxation rate reaches zero.}\label{fig:strong_models}
\end{figure*}

Below $T_N$, soft transverse modes can make spin flips dominate, $R\gg1$, and even at $R=2$ the ratio $a/K$ grows nearly linearly with correlation length, so the spin flip quasiparticle rate can become large while the longitudinal transport strength stays bounded. We follow the linear coefficient $Q=\lim_{\phi_0\to0}\theta/\phi_0$ on NN, NNN and the parabola with $b=1$, for both couplings, correlation lengths $4$ and $16$ and $K=0.003,0.01,0.03$, over $0.1\le a\le10^4$. Once spin flips limit both currents, each decreases approximately as $1/a$, and the ratio approaches a plateau independent of $K$. On the plateau $R=f_ma/K$ exceeds $10^2$, so the plateau is the limit dominated by spin flips. All 36 full curves keep their sign, and Fig.~\ref{fig:strong_models} shows $K=0.01$.

For $\Lambda_A$ at correlation length $4$, the ratio retains $0.47$, $0.53$ and $0.39$ of its band value on NN, NNN and the parabola, and at $16$ these values fall to $0.35$, $0.37$ and $0.25$. For $\Lambda_G$ the corresponding values are $0.35$, $0.41$ and $0.28$ at $4$, and $0.24$, $0.29$ and $0.17$ at $16$. The velocity contribution alone approaches $V/C=-0.500$, $-0.620$ and $1$ at $4$. On NN and the parabola the full plateau is smaller in magnitude than this velocity contribution, and on NNN it is larger. In the limit dominated by spin flips, $P_{\rm drive}/C=-0.222$ and $P_{\rm coll}/C=0.266$ on NN, and $0.116$ and $-0.727$ in the parabola, for $\Lambda_A$ at $4$. The exact solution of Eq.~\eqref{eq:kinetic} explains what sets the remaining ratio. At every scattering strength, the occupation that follows the band shift is $d_{\mathbf k}F'g$, with $g=\Mcal_+^{-1}v_x$ the distribution that carries the charge current, so this part decays only with the charge current [Eq.~\eqref{eq:exactsolution}]. As forward flips come to dominate, they equalize the two spin distributions at equal energy, and $S$ approaches the spin current of the common distribution $Fg$ placed on the two split Fermi lines, with corrections of order $1/r_T$ [Eq.~\eqref{eq:plateaulimit}]. The remaining ratio is set by the bands and by the shape of $g$. The transport time approximation predicts a sign change on NN and in the parabola, and on NNN at $16$ its total local relaxation rate becomes negative or zero at 69 parameter points.

Reaching the plateau in samples with well defined quasiparticles requires $E_F\tau_{\rm bg}/\hbar\gtrsim4\times10^2$ at correlation length $4$, and values up to about $10^4$ at $16$, for $\hbar\Gamma^{\rm tot}_{\rm out}/E_F<0.1$, with the Fermi energy measured from the band bottom. At correlation length $16$ the Born condition $|\mathbf v|/\Gamma_{\rm out}>\xi$ of Sec.~\ref{sec:model}, evaluated at the antinode of the NN band, raises this requirement by a further factor of about $1.4$.

\section{A disorder series measures the slope independent of the order amplitude}\label{sec:temperature}
In a material, temperature changes the splitting, the correlation length and both scattering strengths together, so a temperature scan mixes these effects. A series of samples with different disorder separates them. When disorder changes only the background time, $K$ and $a$ scale with $\tau_{\rm bg}$ at fixed fluctuation strengths, so samples at one temperature follow a line of fixed $R$. With a constant relaxation time, $\theta=\phi_0Q_{\rm CRT}$ is independent of $\tau_{\rm bg}$, so $\theta/\theta_{\rm bg}=Q/Q_{\rm CRT}$, where $\theta_{\rm bg}$ is the spin splitter ratio in samples dominated by background scattering. The slope $s$ then follows from measured ratios alone, independent of $\phi_0$. In $x=\rho_{\rm mag}/\rho_{\rm bg}$, $\rho_{\rm bg}$ is the residual resistivity and $\rho_{\rm mag}$ the magnetic resistivity at the measurement temperature, and phonons enter through Eq.~\eqref{eq:phononbackground} below. Because $\rho_{\rm mag}$ and the phonon resistivity both stay fixed along the series, $\rho_{\rm mag}$ comes from an independent measurement, for example by subtracting the resistivity of an isostructural nonmagnetic reference. To first order $\rho_{\rm mag}$ is independent of $\tau_{\rm bg}$, as in the scaling analysis that separates residual and phonon contributions to the anomalous Hall effect \cite{Tian2009}. The prediction is $1-\theta/\theta_{\rm bg}\simeq s\rho_{\rm mag}/\rho_{\rm bg}$ with $s=1.4$ to $2.4$ for $\Lambda_A$ and up to $5.0$ for $\Lambda_G$ on the reference bands, against $0.3$ to $1.1$ for $\Lambda_A$ and up to $2.4$ for $\Lambda_G$ from the transport time approximation. Since $s=(s_L+Rs_T)/(1+R)$, its sign and size constrain the spin flip fraction of the scattering, which charge transport alone leaves open, once $s_L$ and $s_T$ are taken from a band calculation. Because $s_L$ stays of order one while $s_T$ grows in proportion to $\xi$, a slope that grows in proportion to $\xi$ toward $T_N$ is the signature of spin flip critical fluctuations independent of band details. The calculated series stay within 10 percent of this linear law while $sx\lesssim0.04$ to $0.05$.

The comparison applies to samples with the same $\phi_0$, magnetic spectrum and temperature, with additive resistivities. Samples are best compared at equal $T/T_N$, with the ordered moment checked independently, because disorder can reduce it. The analysis treats the residual background as a single relaxation rate. Phonons scatter anisotropically and can change the ratio by themselves \cite{He2025}, so to first order
\begin{equation}
1-\frac{\theta}{\theta_{\rm bg}}=s\frac{\rho_{\rm mag}}{\rho_{\rm bg}}+s_{\rm ph}\frac{\rho_{\rm ph}}{\rho_{\rm bg}},
\label{eq:phononbackground}
\end{equation}
where $\theta_{\rm bg}$ and $\rho_{\rm bg}$ refer to residual scattering alone. Along a disorder series $\rho_{\rm ph}$ and $\rho_{\rm mag}$ stay fixed while $\rho_{\rm bg}$ changes, so the phonon term shifts the measured slope to $s+s_{\rm ph}\rho_{\rm ph}/\rho_{\rm mag}$ at every disorder level. The measured slope equals $s$ for $|s_{\rm ph}|\rho_{\rm ph}\ll|s|\rho_{\rm mag}$, a stronger condition than $\rho_{\rm ph}\ll\rho_{\rm bg}$, or after an independent determination of $s_{\rm ph}$. For a model anisotropic background, a slope accurate to 10 percent requires $\rho_{\rm ph}/\rho_{\rm mag}$ below about 1 to 3 on the NN band and 3 to 6 on the NNN band (Supplemental Material Sec.~S11). Defects that carry local moments add spin flip scattering and change the fluctuation strengths, so the series uses nonmagnetic defects. A spin splitting torque measures $\theta$ multiplied by a transmission factor $\mathcal T$ that contains the interface and the spin diffusion length, which depends on disorder and on $R$. For $\mathcal T(x)/\mathcal T(0)=1+tx$, the measured slope is $s-t$, so a thickness series at each disorder level fixes $t$.

Reversing the magnetic domain reverses the contribution linear in $\phi_0$, so antisymmetrizing the measured spin conductivity between the two domains selects it as a whole and removes the extrinsic side jump and skew scattering contributions of Ref.~\cite{Sarkar2026}, which are even under time reversal and hence even in $\phi_0$.

The identity $Q=b+P/C$ holds for the NN and parabolic bands. For general bands $V/C$ changes with scattering, by up to about 12 percent in our survey, so the comparison with a measured $Q$ isolates the occupation contribution once $V/C$ comes from the same transport calculation that reproduces the measured resistivity.

\section{Conclusion}\label{sec:conclusion}
Critical magnetic fluctuations treat charge and spin currents differently. Spin conserving fluctuations keep the initial spin decay rate of order the charge decay rate, while nearly forward spin flips equalize the two spin populations at equal energy at the quasiparticle rate, and energy conservation on the split bands leaves the energy shift of the occupation to momentum relaxation at every scattering strength. The slope of the spin splitter ratio at weak scattering interpolates between a bounded spin conserving value and a spin flip value that, in two dimensions, grows nearly linearly with correlation length. Finite splitting blocks the forward flips close to $T_N$, so the growth is best seen in altermagnets whose splitting is small compared with the Fermi energy, and strong flips leave a finite fraction of the lattice band ratio in clean samples, which approaches the spin current of the charge carrying distribution on the two split Fermi lines. On the nearest neighbor square lattice band spin flips drive the reduced ratio to the velocity only value, and in the parabolic band scattering alone moves the ratio away from the band value $b$. Equal resistivity can accompany opposite changes of the spin splitter ratio, which lifetime and transport time approximations miss. A disorder series at fixed reduced temperature measures the slope independent of the order amplitude, and its growth toward $T_N$ identifies spin flip critical fluctuations, once phonon and interface effects are constrained. An injected spin current additionally requires boundary conditions and spatial kinetics \cite{Sigales2026,Zarzuela2025}.

\appendix
\section{Energy conservation and nearly forward spin flips}\label{sec:transverse}
Elasticity makes $F(E)$ constant on each energy shell. In the term of Eq.~\eqref{eq:directM1} that describes scattering into a state, the derivative of the final state occupation $F(E')v_x'$ produces a term proportional to $F'$. Adding it to $\Lcal_-(d_{\mathbf k}F'v_x)$ cancels the spin flip relaxation of the energy shift term and gives
\begin{align}
\Lcal_-X_1+\Lcal_1X_0&=F(\Lcal_-u_x+\Lcal_1v_x)\nonumber\\
&\quad+F'd_{\mathbf k}\Lcal_+v_x,
\label{eq:thermalcancel}
\end{align}
where $\Lcal_1v_x$ is the same derivative operator acting on the bare velocity $v_x$. Inserting Eq.~\eqref{eq:thermalcancel} into Eq.~\eqref{eq:prb_rates} gives
\begin{equation}
\Gamma_sS_0=\langle X_1,\Lcal_+v_x\rangle+\langle Fv_x,\Lcal_-u_x+\Lcal_1v_x\rangle.
\label{eq:thermalrate}
\end{equation}
The occupation change that only shifts the band energy relaxes through the momentum relaxation operator $\Lcal_+$. An electron that flips its spin conserves its energy and lands on the other spin band where the band shift had already placed the occupation, so the flip leaves this part of the spin imbalance in place.

The same identity holds with $v_x$ replaced by any distribution $g$,
$\Lcal_-(d_{\mathbf k}F'g)+\Lcal_1(Fg)=F'd_{\mathbf k}\Lcal_+g+F\Lcal_1g$, where $\Lcal_1g$ acts on the bare $g$. Because the collision operators at zero splitting act within each energy shell, $h_+=\Mcal_+^{-1}X_0=Fg$ with $g=\Mcal_+^{-1}v_x$, the distribution that carries the charge current. Inserting
\begin{equation}
h_-=d_{\mathbf k}F'g+F\Mcal_-^{-1}(u_x-\Mcal_1g)
\label{eq:exactsolution}
\end{equation}
into $\Mcal_-h_-$ and using the identity with $\Mcal_+g=v_x$ reproduces $X_1-\Mcal_1h_+$, so Eq.~\eqref{eq:exactsolution} is the exact solution of Eq.~\eqref{eq:kinetic}, with $\Mcal_1g=\tau_{\rm bg}\Lcal_1g$. The occupation that only shifts the band energy is $d_{\mathbf k}F'g$ at every scattering strength, the energy shift of the charge carrying distribution, and it relaxes only as fast as the charge current does. Equation~\eqref{eq:thermalcancel} is its first order form.

For transverse fluctuations that scatter only with small momentum transfer, $q\ll\bar k_F$, with a rate $\Gamma_T$ of scattering out of a state, the leading terms are $\Lcal_{+,T}v_x\to0$, $\Lcal_{-,T}u_x\to2\Gamma_Tu_x$ and $\Lcal_{1,T}v_x\to-2d_{\mathbf k}\Gamma_T\partial_Ev_x$, where $\partial_E=(\mathbf v\cdot\nabla_{\mathbf k})/(\hbar|\mathbf v|^2)$. Thus
\begin{equation}
\Gamma_{s,T}S_0\simeq2\int_{\mathbf k}F\Gamma_Tv_x(u_x-d_{\mathbf k}\partial_Ev_x).
\label{eq:forwardspin}
\end{equation}
Here $u_x-d_{\mathbf k}\partial_Ev_x$ is the change of $v_x$ per unit $\sigma\phi_0$ at fixed energy, because the spin $\sigma$ line at energy $E$ lies a distance $-\sigma\phi_0d_{\mathbf k}/(\hbar|\mathbf v|)$ from the unsplit line along its normal. Each spin drifts with its own velocity, so with background scattering alone $F(u_x-d_{\mathbf k}\partial_Ev_x)=h_--d_{\mathbf k}\partial_Eh_+$ is the imbalance of the two spin populations at equal energy, and Eq.~\eqref{eq:forwardspin} states that forward flips remove this imbalance at the rate $2\Gamma_T$. The initial spin decay rate is of order the quasiparticle scattering rate, while the initial charge decay rate is of order the transport rate. Using $\Gamma_{T,\rm tr}(0)/2$ for the charge decay rate gives $s_T\simeq\alpha r_T+O(1)$, with
\begin{equation}
\alpha=\frac{4\int_{\mathbf k}F\Gamma_Tv_x(u_x-d_{\mathbf k}\partial_Ev_x)}{S_0\Gamma^{\rm avg}_{T,\rm out}(0)}.
\label{eq:forwardalpha}
\end{equation}
The averages in $r_T$ are evaluated on the Fermi line. At finite temperature, the exact charge rate is thermally averaged, so the coefficient of $r_T$ also contains the factor $\Gamma_{T,\rm tr}(0)/(2\Gamma_{c,T})$.

For the Ornstein--Zernike form of the fluctuations in this limit, $\Gamma_T\propto\Phi_d^2/|\mathbf v|$ for both couplings. Equation~\eqref{eq:forwardalpha} then gives $2.65$ for the NN band and $2.32$ for the NNN band, and the two couplings share these values because they coincide at zero momentum transfer. The slopes calculated between $\xi\bar k_F=32$ and 64 are about $2.62$ and $2.21$, and the thermal factor above accounts for most of the remaining difference. For the parabolic band, the value at zero temperature is $5/2$, and at $k_BT/\mu_0=0.05$ the thermal band integral gives $2.54$.

The scaling with correlation length follows directly from the forward kernel. Parametrize a short part of the Fermi line by the transfer $q$ along it. The integral $\int dq/(\xi^{-2}+q^2)$ grows as $\pi\xi$, whereas the transport weight is quadratic in $q$ and makes $\int dq\,q^2/(\xi^{-2}+q^2)$ approach a finite value set by the band. Thus $r_T\propto\xi\bar k_F$ at long correlation length. Fits over $\xi\bar k_F=8$, $16$ and $32$ give exponents $0.93$ to $0.98$ for $r_T$ and $1.02$ to $1.06$ for $s_T$. On a three dimensional Fermi surface the transfer has two components along the surface, and $\int d^2q\,(\xi^{-2}+q^2)^{-1}\simeq2\pi\ln(\xi q_c)$, with $q_c$ set by the band, while the transport integral stays finite, so $r_T\propto\ln(\xi\bar k_F)$.

When spin flips limit both currents, $\tau_{\rm bg}\Lcal_\pm\gg1$, Eq.~\eqref{eq:exactsolution} makes $S$ and $C$ both scale as $1/a$, so their ratio depends only on the bands and on the shape of the scattering. In the forward limit, $\Lcal_{-,T}\to2\Gamma_T$ and $\Lcal_{1,T}g\to-2d_{\mathbf k}\Gamma_T\partial_Eg$ for a smooth $g$, so the last term of Eq.~\eqref{eq:exactsolution} approaches $Fd_{\mathbf k}\partial_Eg$ up to corrections of order $1/r_T$, and
\begin{equation}
S\to\int_{\mathbf k}\left[u_xFg+v_xd_{\mathbf k}\partial_E(Fg)\right].
\label{eq:plateaulimit}
\end{equation}
In this limit $h_--d_{\mathbf k}\partial_Eh_+\to0$, so the two spin populations are equal at equal energy, and Eq.~\eqref{eq:plateaulimit} is the spin current of a common distribution $Fg$ placed on the two split Fermi lines, carried entirely by their velocity difference.

%

\clearpage
\setcounter{section}{0}
\setcounter{subsection}{0}
\setcounter{figure}{0}
\setcounter{table}{0}
\setcounter{equation}{0}

\makeatletter
\def\@hangfrom@section#1#2#3{\@hangfrom{#1#2}\MakeTextUppercase{#3}}
\let\@sectioncntformat\@seccntformat
\makeatother
\counterwithout*{equation}{section}
\setcounter{secnumdepth}{1}
\renewcommand{\thesection}{S\arabic{section}}
\renewcommand{\thesubsection}{S\arabic{section}.\arabic{subsection}}
\renewcommand{\thefigure}{S\arabic{figure}}
\renewcommand{\thetable}{S\arabic{table}}
\renewcommand{\theequation}{S\arabic{equation}}
\providecommand{\theHsection}{}
\renewcommand{\theHsection}{SM.\arabic{section}}
\providecommand{\theHsubsection}{}
\renewcommand{\theHsubsection}{SM.\arabic{section}.\arabic{subsection}}
\providecommand{\theHfigure}{}
\renewcommand{\theHfigure}{SM.\arabic{figure}}
\providecommand{\theHtable}{}
\renewcommand{\theHtable}{SM.\arabic{table}}
\providecommand{\theHequation}{}
\renewcommand{\theHequation}{SM.\arabic{equation}}
\onecolumngrid
\begin{center}
{\large\bfseries Supplemental Material for ``Spin-conserving and spin-flip magnetic scattering in the altermagnetic spin-splitter effect''\par}
\vspace{0.8em}
Yunxi Jiang and Hao Yu
\end{center}

\twocolumngrid

$V$ and $P$ are the velocity and occupation contributions to the spin current, $S=V+P$ is the total spin response and $C$ is the charge response. With electron charge $-e$, background time $\tau_{\rm bg}$ and dimensionless order amplitude $\phi_0$, the conductivities are $\sigma^s=2e^2\tau_{\rm bg}\phi_0S$ and $\sigma^c=2e^2\tau_{\rm bg}C$. Lattice energies are in units of $t_1$, with $\hbar=k_B=1$, and collision matrices are in units of $\tau_{\rm bg}^{-1}$. NN denotes the band with only nearest neighbor hopping and NNN the band that also includes next nearest neighbor hopping.

\section{Closure on the two spin occupations}\label{sec:sm_closure}
This section supports Sec.~II~A of the main text. For the coupling $H_{\rm int}=\sum_\alpha\Lambda c^\dagger\sigma_\alpha cB_\alpha$ between electrons and magnetic fluctuations $B_\alpha$, we take fluctuations with zero mean and short memory, with correlators $C_{xx}=C_{yy}$ and $C_{xz}=C_{zx}=C_{yz}=C_{zy}=0$. The closure needs only the vanishing cross correlations between longitudinal and transverse fluctuations, and $C_{xx}=C_{yy}$ expresses the assumed axial symmetry. An antisymmetric part $C_{xy}=-C_{yx}$ also preserves diagonal occupations. For classical static fluctuations in a centrosymmetric magnet, inversion and complex conjugation make the correlator real and symmetric, and the quantum antisymmetric part is smaller by $\hbar\omega_q/k_BT$ and vanishes in the quasistatic limit. Conjugation by $\sigma_z$ preserves diagonal occupations, while conjugation by $\sigma_x$ or $\sigma_y$ exchanges them. The Born gain and loss terms keep $\operatorname{span}\{I,\sigma_z\}$ closed for a drive diagonal in spin, and the occupations of the two spin states describe the response to the electric field completely. This conserving-collision construction is consistent with the framework of Refs.~\cite{SMBaymKadanoff1961,SMBaym1962}.

\section{Model, density and calibration of the scattering strengths}\label{sec:sm_model}
This section supports Sec.~II~B of the main text. In lattice units, the unsplit dispersion, the exchange energy per unit order amplitude and the two spin bands are
\begin{align}
\epsilon_0(\kk)&=-2(\cos k_x+\cos k_y)-4t_2\cos k_x\cos k_y,\nonumber\\
d_\kk&=\cos k_x-\cos k_y,\qquad
\varepsilon_{\kk\sigma}=\epsilon_0(\kk)-\mu+\sigma\phi_0 d_\kk,
\label{eq:lattice}
\end{align}
with momentum in inverse lattice spacings and $\sigma=\pm1$ along a fixed axis. The velocities are $v_i=\partial_{k_i}\epsilon_0$ and $u_i=\partial_{k_i}d_\kk$.

We keep the density equal to its value at zero temperature and zero splitting. With $n(\mu,\phi_0,T)$ the total density of both spins, the chemical potential solves $n(\mu(\phi_0,T),\phi_0,T)=n(\mu_0,0,0)$, with $\mu_0/t_1=-1$ for the NN band and $\mu_0/t_1=-1.2$ for the NNN band. Exchanging the spin labels maps $\phi_0$ to $-\phi_0$ and leaves the total density unchanged, so $\partial\mu/\partial\phi_0=0$ at $\phi_0=0$. The Fermi--Dirac function is $f(E)$, its thermal weight is $F(E)=-df/dE$ and $F'=dF/dE$. With background scattering alone, at $k_BT/t_1=0.05$, the NN band has $C^V=-0.182$, $C^P=-0.169$ and $C_0=0.363$, and the NNN band has $C^V=-0.181$, $C^P=-0.323$ and $C_0=0.300$. The velocity and occupation contributions thus have the same sign and add up.

On the energy contour $\varepsilon_{\mathbf k\sigma}=E$, the quadrature weight $w_i$ of node $i$ contains the density of states, and the state index runs over both spin contours. In continuum notation, $w_i$ approximates $d\ell/[(2\pi)^2\hbar|\mathbf v_\sigma|]$. With the symmetric scattering amplitude $A_{ij}$ between nodes we form $G_{ij}=w_iA_{ij}w_j$ and the symmetrized collision matrix
\[
L=w^{-1/2}[\operatorname{diag}(G\mathbf 1)-G]w^{-1/2},
\]
where $w$ is the diagonal matrix of weights and $\mathbf1$ the constant vector. The conserved number vector is $\sqrt w$, with $L\sqrt w=0$. For an occupation displacement $\boldsymbol\psi$, the quadratic form is $\frac12\sum_{ij}G_{ij}(\psi_i-\psi_j)^2$, which is nonnegative and vanishes on a connected component only for a constant occupation. The scattering amplitude is proportional to $\Lambda_m^2/(\xi^{-2}+q_{\rm lat}^2)$ with the couplings $\Lambda_A$ and $\Lambda_G$ of the main text. Longitudinal transitions preserve spin and transverse transitions exchange it, and each uses the actual final contour of that spin.

For the reference charge drive, the longitudinal rate is $\Gamma_{L,\rm tr}=2\mathbf b^TL_L\mathbf b/(\mathbf b^T\mathbf b)$ with $b_i=\sqrt{w_i}v_{x,i}$, where $L_L$ is the symmetrized longitudinal matrix, and the spin conserving strength is $K=\Gamma_{L,\rm tr}\tau_{\rm bg}$. The factor of two makes the charge response on this contour fall with initial slope $-1/2$ in $K$, so purely longitudinal scattering gives the relative magnetic resistivity $K/2$ at weak scattering. The rate $\Gamma_{T,\rm tr}$ in $r_T$ of the main text is defined in the same way for transverse scattering, and both are twice the average of the transport time rate $\gamma_{\rm tr}$ of Eq.~(19) of the main text over the unsplit Fermi line with the weight $v_x^2$. The spin flip strength is $a=2\tau_{\rm bg}\Gamma^{\rm AN}_{T,\rm out}$, with $\Gamma^{\rm AN}_{T,\rm out}$ the rate at which transverse fluctuations scatter electrons out of the states at an antinode. Both strengths are calibrated once on the unsplit reference contour at $E=\phi_0=0$ and kept fixed at every $\pm\phi_0$. The electric drive is odd under inversion, and we solve in this symmetry subspace, using the additional reflection symmetry for the $xx$ response. The conserved number mode lies in the complementary even subspace, and on the driven subspace the background contributes the identity, so $\widetilde{\mathcal M}=I+\tau_{\rm bg}L$ is positive definite.

The same reference moments convert the two strengths into the ratio of fluctuation strengths,
\begin{equation}
 R=\frac{A_T}{A_L}=\frac{a}{K}\frac{t_L}{2o_T}\equiv f_m\frac aK,\qquad
 t_L=\frac{2\langle v_x,L_L^{\rm unit}v_x\rangle_w}{\langle v_x,v_x\rangle_w},
\label{eq:sm_R}
\end{equation}
where $o_T$ is the transverse rate of scattering out of the antinode at unit strength. The factors $f_m$ are $0.150$ and $0.0819$ for the NN band, $0.175$ and $0.116$ for the NNN band, and $0.0696$ and $0.0307$ for the parabolic band, for $\Lambda_A$ and $\Lambda_G$. For example, $a=100$ and $K=0.1$ on the NN band give $R\simeq150$ for $\Lambda_A$ and $82$ for $\Lambda_G$. The two kinds of fluctuations share the same momentum dependence, so their contributions to charge transport differ only by their strengths. Rates of scattering out of nodal states are finite, because the coupling also depends on the final momentum.

For the quasiparticle condition, we use $E_F=\mu(T)-\epsilon_{\min}$ and $X=E_F\tau_{\rm bg}/\hbar$. With a constant background, $\tau_{\rm bg}\Gamma_{\rm out}^{\rm tot}=1+\tau_{\rm bg}(\Gamma_{L,\rm out}+\Gamma_{T,\rm out})$, so the condition $\hbar\Gamma_{\rm out}^{\rm tot}/E_F<\eta$ requires $X>\tau_{\rm bg}\Gamma_{\rm out}^{\rm tot}/\eta$. For the NN band at $K=0.1$ and $\eta=0.1$, the bounds at the antinode are about 18 and 21 at $a=1$, 163 and 166 at $a=30$, and 513 and 516 at $a=100$, for $\Lambda_A$ and $\Lambda_G$.

\section{Energy contour response and saddle integration}\label{sec:sm_contour}
This section supports Sec.~II~C of the main text. We evaluate the kinetic equation of the main text,
\begin{equation}
h_+=\Mcal_+^{-1}X_0,\qquad
h_-=\Mcal_-^{-1}(X_1-\Mcal_1h_+),
\label{eq:fixedresponse}
\end{equation}
with $X_0=Fv_x$ and $X_1=Fu_x+d_\kk F'v_x$, on the combined spin contours at fixed energy. The physical coefficients are $V=\int_\kk u_xh_+$ and $P=\int_\kk v_xh_-=S-V$. We write $y_i=\sqrt{w_i}\,\delta f_i/[-eE_x\tau_{\rm bg}F(E)]$ and $b_i=\sqrt{w_i}v_{x,i}$, so the contour equation is $\widetilde{\mathcal M}\mathbf y=\mathbf b$, where the tilde denotes the symmetrized representation. Let $\mathbf j$ be the spin current vector on the combined contours, with components $j_{\sigma,i}=(\sigma/2)\sqrt{w_{\sigma,i}}v_{x,\sigma,i}$ that include the average over the two spins. The spin conductivity is
\[
\frac{\sigma^s_{xx}(\phi_0)}{2e^2\tau_{\rm bg}}=
\int dE\,F(E)\,\mathbf j(\phi_0)^T\mathbf y(\phi_0).
\]
With subscripts 0 and 1 for the value and the derivative at $\phi_0=0$, differentiating the contour equation gives $\mathbf y_1=\widetilde{\mathcal M}_0^{-1}(\mathbf b_1-\widetilde{\mathcal M}_1\mathbf y_0)$, and the total coefficient is
\[
S=\int dE\,F(E)
\left[\mathbf j_1^T\mathbf y_0+
\mathbf j_0^T\widetilde{\mathcal M}_0^{-1}\mathbf b_1-
\mathbf j_0^T\widetilde{\mathcal M}_0^{-1}
\widetilde{\mathcal M}_1\mathbf y_0\right].
\]
The three terms describe the change of the measured current, the change of the drive and the change of the collisions. The contour positions, weights, velocities and scattering amplitudes are all differentiated.

\subsection{Thermal integration near a saddle}\label{sec:sm_thermal}
The energy range is divided at the band edges $-4-4t_2-\mu$ and $4-4t_2-\mu$ and at the spin dependent saddle energies $4t_2-\mu\pm2\phi_0$. On each interval $[E_-,E_+]$ we integrate in $\theta\in(0,\pi/2)$ with $E=E_-+(E_+-E_-)\sin^2\theta$, and each spin branch contributes on its own interval. The thermal weight is $F(E)=f(E)[1-f(E)]/T$.

Near a saddle, the derivative of the thermal integrand has large contributions of opposite sign on the two sides of the saddle energy. We add them at matched offsets before integrating over the offset. With $E_c$ the unsplit saddle energy, $g(E)$ the thermal integrand of $S$ and $r_E$ the radius of the paired interval, the exact rearrangement is
\begin{equation}
\int_{E_c-r_E}^{E_c+r_E}g(E)dE
=\int_0^{r_E}[g(E_c-\delta)+g(E_c+\delta)]d\delta.
\label{eq:paired}
\end{equation}
Positive offsets are integrated in logarithmically spaced cells, and the innermost interval is kept with a midpoint contribution.

The parabolic band uses the dispersion and exchange energy of the main text with $k_F^2=2m\mu_0/\hbar^2$ and units $m=d_0=\mu_0=\hbar=1$. In polar coordinates $\mathbf k=r(\cos\varphi,\sin\varphi)$, the spin bands and the contour measure $d\Omega_\sigma$ are
\begin{align*}
\varepsilon_\sigma&=\frac{\hbar^2r^2}{2m}
\left[1+\sigma\phi_0\frac{d_0}{\mu_0}\cos2\varphi\right]-\mu,\\
d\Omega_\sigma&=\frac{m\,d\varphi}
{(2\pi)^2\hbar^2[1+\sigma\phi_0(d_0/\mu_0)\cos2\varphi]},
\end{align*}
and the density is
\begin{equation}
n(\mu,\phi_0,T)=\frac{T\ln(1+e^{\mu/T})}{\pi\sqrt{1-\phi_0^2}}.
\label{eq:continuumdensity}
\end{equation}
The coupling $\Lambda_G$ multiplies the endpoint average by $e^{-q^2/8}$ with momenta in these units, and the bare occupation coefficient vanishes analytically.

\section{Slope over a wide range of band parameters}\label{sec:sm_survey}
This section supports Sec.~II~D of the main text. The slope at weak scattering is evaluated on 72 lattice cells for the two couplings, $R=1,2$ and $\xi\bar k_F=2,4,8$, giving 864 parameter sets. A Latin hypercube sample adds 2000 lattice sets with $-0.35\le t_2/t_1\le0.15$, $0.10\le\nu\le0.75$, $1.5\le\xi\bar k_F\le10$ and $0.25\le R\le8$, keeping $|\mu-4t_2|\ge3k_BT$. Correlation length and $R$ are sampled logarithmically, at temperatures $k_BT/t_1=0.02,0.05,0.1$. Of these, 1334 use the $d$-wave couplings and 666 a constant coupling, which represents the coupling to fluctuations of the uniform magnetization. A parabolic sample covers the same intervals of correlation length and $R$ at $b=1$, with 200 points per coupling.

\begin{figure}[tbp]\centering\includegraphics[width=3.4in]{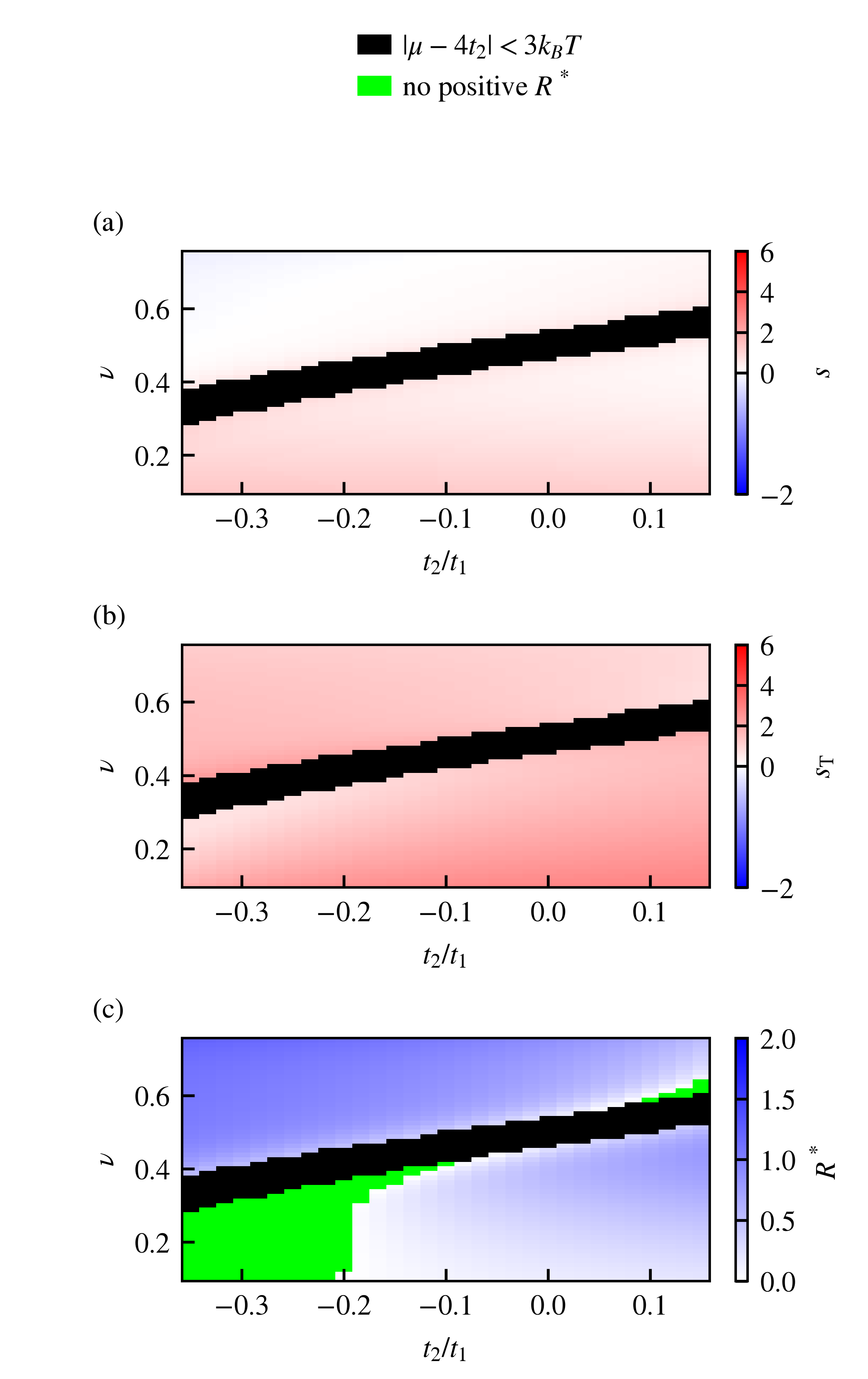}
\caption{Slopes at weak scattering and the crossover between enhancement and suppression for $\Lambda_A$, $\xi\bar k_F=2$ and $k_BT/t_1=0.05$, across hopping ratio $t_2/t_1$ and filling $\nu$. (a) Total slope $s$ at $R=1$. (b) Pure spin flip slope $s_T$. (c) Positive crossover ratio $R^*=-s_L/s_T$. Black cells lie within $3k_BT$ of the saddle energy $4t_2$. Green cells in (c) have $s_L$ and $s_T$ of the same sign. White denotes zero on each color scale. The scales saturate outside $-2\le s,s_T\le6$ and $0\le R^*\le2$.}\label{fig:sm_signslice}\end{figure}

On the lattice grid the median slope is 2.38, and 8 of the 864 slopes are negative or zero, the smallest $-0.06$. In the wider lattice sample the median is 2.08, with 62 of 1334 slopes negative or zero, the smallest $-0.93$. The constant coupling gives a median of 0.99, with 13 of 666 slopes negative or zero. All 600 parabolic slopes are positive. Every transverse limit $s_T$ is positive, so every negative slope has $s_L<0$ and $R\le-s_L/s_T$. All eight negative grid slopes lie above the van Hove saddle, at $\Lambda_A$, $R=1$, $\xi\bar k_F=2$, $t_2/t_1=-0.3,-0.2$ and $\nu=0.6,0.7$, and 48 of the 62 in the wider sample lie above it as well. Figure~\ref{fig:sm_signslice} maps this slice at $k_BT/t_1=0.05$. On the slice $R=2$, $\xi\bar k_F=4$, $k_BT/t_1=0.05$, all 2852 slopes outside the saddle window are positive.

\section{Spin conserving and spin flip limits}\label{sec:sm_channels}
This section supports Secs.~II~D and III~C of the main text. At zero splitting and fixed $A_L+A_T$, $\Lcal_+$ and the spin averaged scattering rates are independent of $R$, while the spin block depends on it, and $s=(s_L+Rs_T)/(1+R)$ holds throughout the parameter designs of Sec.~\ref{sec:sm_survey}. At $k_BT/t_1=0.05$, the spin conserving and spin flip limits are $s_L=-0.50$ and $s_T=3.77$ on the NN band, $s_L=0.38$ and $s_T=2.37$ on the NNN band, and $s_L=0.20$ and $s_T=8.41$ in the parabolic band, for $\Lambda_A$. For $\Lambda_G$ they are $-0.056$ and $7.53$, $0.74$ and $3.92$, and $2.40$ and $20.29$. The spin flip limit is large on every band, and almost all of it comes from the change of the collision process with the order. The spin conserving part of that change stays below 0.11 on both lattice bands, because $d_\kk-d_{\kk'}$ vanishes for forward scattering.

\subsection{Unequal correlation lengths}
In the ordered phase, the transverse static susceptibility is of Goldstone type and can have a longer correlation length than the longitudinal one. We fix $\xi_L\bar k_F=4$ and vary $\xi_T\bar k_F$ from 4 to 64. The two kinds of fluctuations then relax the charge current at different rates, so charge transport depends on $R$ as well as on the total strength. At fixed $R$ the slope grows roughly in proportion to $\xi_T$, and at fixed $a/K$ much more slowly.

\section{Comparison of kinetic models}\label{sec:sm_hierarchy}
This section supports Secs.~III~A and III~D of the main text. The comparison covers 62 configurations, all at $k_BT/t_1=0.05$ and $\xi\bar k_F=4$, or $\xi k_F=4$ in the parabolic band. They comprise 40 points on the lines $R=1,2$ for both lattice bands and both couplings at $K=0.03,0.1,0.3,1,3$, 12 NN points at $K=0.1$ and $a=0.3,1,3,10,30,100$, and ten parabolic points at $K=0.1$ and $a=0.1,0.3,1,3,10$. The full calculation decreases $|Q|$ at all 62 points, by 3.3 to 62 percent, and keeps its sign. For each model $j$ we write $\Delta_j=Q_j-Q_{\rm CRT}$ and compare it with the full correction through $f_j=\Delta_j/\Delta_{\rm full}$, at the 59 points where $|C\Delta_{\rm full}|\ge0.05C_0$.

The first order response is linear in $\Mcal_1$ at fixed inverse operators, so the full correction splits into three parts,
\begin{equation}
 \Delta_{\rm full}=\Delta_{0}+(\Delta_{\rm out}-\Delta_{0})
 +(\Delta_{\rm full}-\Delta_{\rm out}),
\label{eq:hierarchy_split}
\end{equation}
namely the unsplit collision result, the change of scattering out of a state and the change of scattering into a state. The fraction $1-f_0$ carried by the change of the collision process with the order exceeds one, so that the unsplit collision model and the full calculation correct $Q$ in opposite directions, at ten of the 30 NN points, all with $K\le0.3$ and $a\le2$, at all nine parabolic points and at none of the 20 NNN points. On the lattice at $K\ge0.3$, the change of scattering out of a state supplies 88 to 127 percent of the full correction. In the parabolic band at $a=3$, this part alone gives $Q=0.986$ and $1.05$ for $\Lambda_A$ and $\Lambda_G$, against the full $0.675$ and $0.596$, so the change of scattering into a state produces the suppression.

\subsection{Lifetime and transport time approximations}
The two approximations correspond to the self-energy and momentum-relaxation-time approaches used in charge transport \cite{SMPonce2020,SMClaes2022}.
The correction of the lifetime approximation has the wrong sign at 21 of the 30 NN points and at all nine parabolic points, and it reaches at most 42 percent of a correction with the correct sign. The transport time approximation overestimates the correction by up to a factor of 4.3, and its correction has the wrong sign at four of the 59 points. At weak scattering, the transport time slopes of $Q$ are 0.52 and 1.08 for NN $\Lambda_A$ at $R=1,2$, against the full 1.64 and 2.35, and 0.33 and 0.51 for NNN $\Lambda_A$, against 1.38 and 1.71. Collisions computed on the unsplit bands give the opposite sign, $-1.57$ and $-1.95$ on the NN band. At strong scattering, the transport time approximation predicts a reversal of $Q$ at nine configurations where the full result keeps its sign, six NN, one NNN and two parabolic. Its local factor $1-\mathbf v'\cdot\mathbf v/|\mathbf v|^2$ is negative at some states, while its integrated rate stays positive and $1+\tau_{\rm bg}\gamma_{\rm tr}$ exceeds unity at every state for all nine.

\subsection{Parabolic band and the form of the coupling}
To see how the parabolic result depends on the coupling, we keep the parabolic band fixed and compare radial, angle only and energy normalized coupling forms, each with the endpoint average and with the factor $e^{-q^2/8}$. Energy normalization removes the radial growth of the form factor, while the Ornstein--Zernike momentum dependence still changes with energy. The full $Q$ lies below its bare value at all thirty points. The ordering $Q_{\Lcal_1=0}>Q_{\rm CRT}$ holds at $a=3$ for all six coupling forms, and it reverses for the angle only and energy normalized couplings at $a=0.1$ and for their versions with $e^{-q^2/8}$ at $a=0.3$. For $\Lambda_G$ at $a=3$, the full calculation gives $Q=0.596$, the transport time approximation $0.613$, the lifetime approximation $1.50$ and the unsplit collision model $1.31$.

Relaxation that depends on angle or energy lifts the parabolic cancellation already with $\Mcal_1=0$. With $d_\kk=b\epsilon\cos2\varphi$ in polar coordinates, define $\mathcal R_{mn}(\epsilon)=\langle\cos m\varphi,\widetilde\Mcal_-^{-1}(\epsilon)\cos n\varphi\rangle$ with angular measure $d\varphi/(2\pi)$ and $\epsilon=\hbar^2k^2/(2m)$. Since $\cos2\varphi\cos\varphi=(\cos\varphi+\cos3\varphi)/2$, the field driven part of the occupation contribution is $P_{\rm drive}=(b/\pi\hbar^2)\int_0^\infty d\epsilon\,[\epsilon F\mathcal R_{11}+(\epsilon^2/2)F'(\mathcal R_{11}+\mathcal R_{13})]$, and integrating the second term by parts gives
\begin{align}
 P_{\rm drive}=-\frac{b}{\pi\hbar^2}\int_0^\infty d\epsilon\,F\bigg[\epsilon\mathcal R_{13}
 +\frac{\epsilon^2}{2}\partial_\epsilon(\mathcal R_{11}+\mathcal R_{13})\bigg].
 \label{eq:harmonicdrive}
\end{align}
The first term describes mixing of the first and third angular harmonics, and the second describes relaxation that depends on energy. Both vanish for the bare inverse operator, for which $\mathcal R_{11}=1/2$ and $\mathcal R_{13}=0$. A scalar angular lifetime $\tau/\tau_{\rm bg}=\bar\tau+\alpha\cos4\varphi$ gives $P/C=-b\alpha/(2\bar\tau)$. At $a=3$ the energy derivative term is the larger of the two for both couplings. For $\Lambda_G$ at small $a$ the mixing term is negative while the energy term is positive, and their competition places a zero of the field driven part near $a=0.061$. At $K=0.1$ and $a=3$ the collision part is larger, with $P_{\rm drive}=0.048$ and $P_{\rm coll}=-0.093$ for $\Lambda_A$, and $0.046$ and $-0.105$ for $\Lambda_G$.

\section{Field driven and collision parts of the occupation contribution}\label{sec:physicaldrive}
This section supports Sec.~III~B of the main text. The field driven part is evaluated in the contour representation of Sec.~\ref{sec:sm_contour}. With the thermal factor divided out of the state, the analytic source pulled back to the unsplit nodes is
\[
 (X_{1,\rm pull})_{\sigma,i}
 =\sigma\sqrt{w_i}\left[u_{x,i}+d_i\frac{F'}{F}v_{x,i}\right],
\]
where the spin sign belongs to the combined spin vector, and
\begin{align}
 P_{\rm drive}&=\int dE\,F\,
 \mathbf j_0^T\widetilde{\mathcal M}_0^{-1}X_{1,\rm pull},\\
 P_{\rm coll}&=S-V-P_{\rm drive},
\end{align}
where the second line uses the complete contour derivative of $S$. Comparing contour quantities with those at fixed momentum requires a connection, because the contour nodes, weights and measurement vectors move with $\phi_0$. For a change of representation $U$, the connection $\Omega=U_1U_0^{-1}$ gives $j_{1,\rm phys}=j_1+\Omega^Tj_0$, whereas $X_{1,\rm pull}=b_1-\Omega b_0$ and $M_{1,\rm pull}=\widetilde M_1-[\Omega,\widetilde M_0]$. The signs differ because sources transform with $U$ and measurements with $U^{-T}$. In the parabolic band this connection separates the longitudinal and transverse contributions to $P_{\rm coll}$, and on the lattice the decomposition uses the analytic drive and the total current.

On the NN band, $P_{\rm drive}$ stays negative at all sampled strengths $a=0,0.03,0.1,0.3,1,3,10,30,100$ for both couplings, while $P$ changes sign at $a\simeq32.75$ for $\Lambda_A$ and $10.82$ for $\Lambda_G$. The field driven part alone keeps the bare sign of the occupation contribution, and the reversal comes from the collision part. At $a=100$ for $\Lambda_A$, $P_{\rm drive}=-0.0113$ and $P_{\rm coll}=0.0127$. In the parabolic band, $P_{\rm drive}$ for $\Lambda_G$ changes sign near $a=0.061$, while the total occupation term stays negative at all sampled points through $a=30$.

\subsection{Sign of the field driven part}
When $u_x=bv_x$, the field driven part is $P_{\rm drive}=bA+B$ with
\begin{equation}
A=\int dE\,F\langle\sqrt wv_x,\widetilde M_-^{-1}\sqrt wv_x\rangle .
\end{equation}
Positive scattering rates that obey detailed balance give $\widetilde M_-\succeq I$ on the odd driven subspace, so $A>0$ at all rates wherever the velocity is finite, and $bA$ always has the sign of the band ratio $b$. $B$ is the response to the energy shift source $d_\kk F'v_x$. With background scattering alone, $A=C_0$ and $B=C^P-C^V$, which gives $A=0.363$ and $B=0.013$ on the NN band at $k_BT/t_1=0.05$ and $A=-B=1/(2\pi)$ in the parabolic band, the exact cancellation of Sec.~III~D of the main text. At the 18 NN reference points through $a=100$, $|B/(bA)|\le0.330$, so the field driven part keeps the bare sign.

\subsection{Varying the spin conserving strength on the lattice}
To determine whether spin flip scattering alone supports the reversal, we vary $K=0,0.03,0.1,0.3,1$ and $a=0,1,3,10,30,100,300$ at the NN reference density for both couplings, keeping the analytic drive and the complete collision derivative. At $K=0.1$ the zeros of $P$ lie near $a=32.8$ for $\Lambda_A$ and $10.8$ for $\Lambda_G$. Setting $K=0$ removes the spin conserving contribution to the collision derivative and leaves reversals near $a=30.9$ and $10.9$, and at $K=1$ the zeros move to $a=49.9$ and $13.1$. Spin flip scattering alone suffices for the reversal. The ratios of fluctuation strengths at the zeros are $R\approx49$ and $8.9$ at $K=0.1$, or $7.5$ and $1.1$ at $K=1$. The corresponding relative magnetic resistivities at $K=0.1$ are about 2.2 and 0.43, and the quasiparticle condition requires $E_F\tau_{\rm bg}/\hbar$ of roughly 70 to 170.

\begin{figure}[tbp]
\centering\includegraphics[width=3.4in]{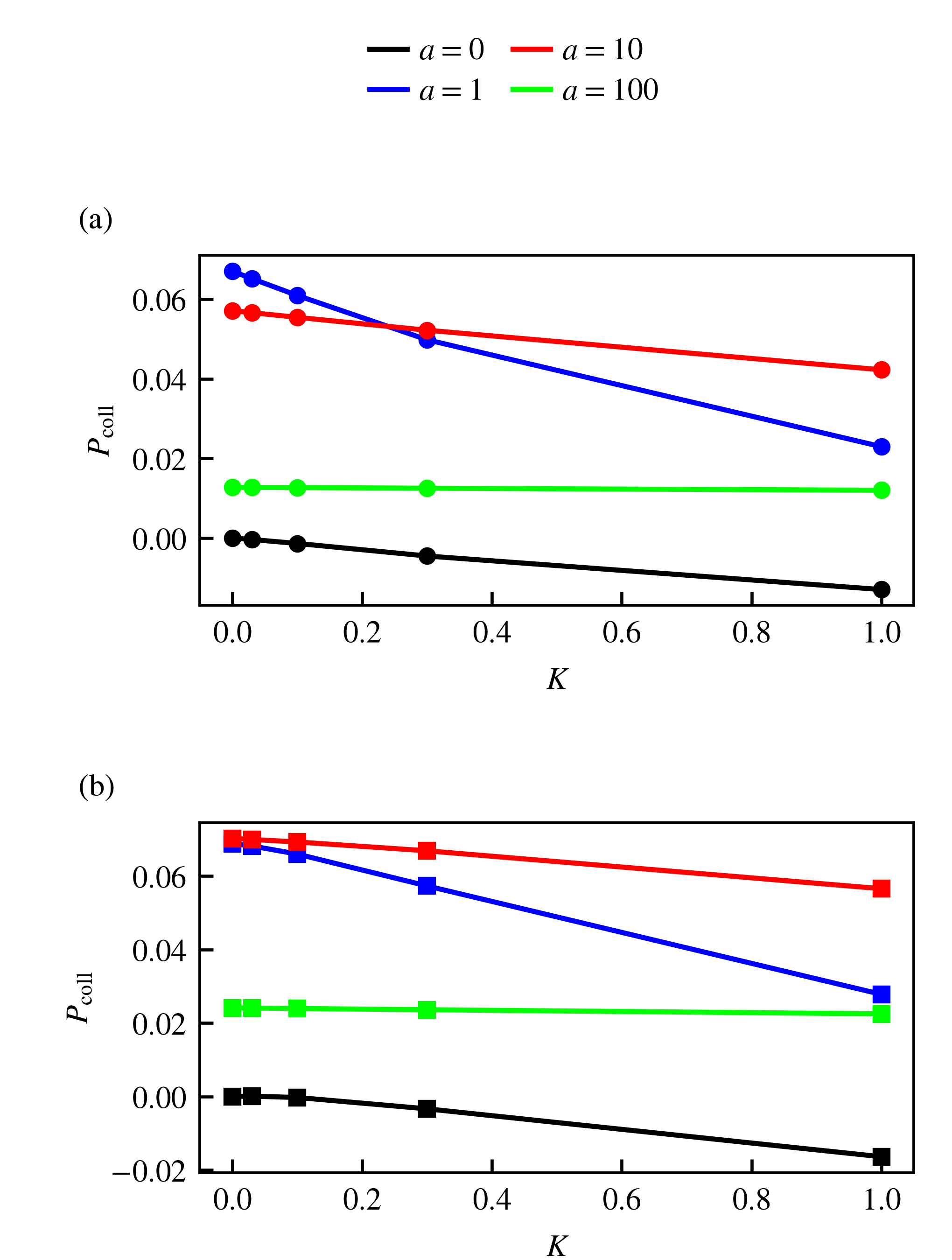}
\caption{Collision contribution $P_{\rm coll}=P-P_{\rm drive}$ in the momentum partition on the NN reference band, as a function of spin conserving strength $K$. Density is fixed, $k_BT/t_1=0.05$ and $\xi\bar k_F=4$. Panels (a) and (b) use $\Lambda_A$ and $\Lambda_G$. Black, blue, red and green denote spin flip strengths $a=0,1,10,100$. The coefficients are shown without normalization by the charge response.}
\label{fig:channel}
\end{figure}

Figure~\ref{fig:channel} shows $P_{\rm coll}$ against $K$. At $a=1$, $P_{\rm coll}$ varies with $K$ by 83 and 71 percent of its mean for the two couplings, and at $a=100$ by only 6 and 7 percent, so the collision part becomes nearly independent of $K$ when spin flip scattering is strong. At small $a$, $P_{\rm coll}(a)-P_{\rm coll}(0)$ grows linearly with $a$, and at $K>0$ the finite value at $a=0$ comes from spin conserving scattering alone.

\subsection{Spin conserving and spin flip parts in the parabolic band}
In the parabolic band the connection gives $M_{1,\rm pull}$ and its spin conserving and spin flip parts. At $a=3$, the spin flip part of $P_{\rm coll}$ is $-0.093$ for $\Lambda_A$ and $-0.105$ for $\Lambda_G$, and the spin conserving part is below $10^{-3}$ in magnitude. The spin flip part is larger throughout the sampled continuum grid, by a factor of at least $1.33$, reached for $\Lambda_G$ at $a=0.05$. At fixed normalized couplings, density and calibrated rates, changing $b=d_0/\mu_0$ multiplies every spin coefficient linear in $\phi_0$ by $b$ and leaves $C$ unchanged.

\section{Comparison of the momentum and energy partitions}\label{sec:sm_zeros}
This section supports Sec.~III~B of the main text. With background scattering alone, the measurement and source derivatives are equal in the symmetric representation at equal energy, so the occupation and velocity terms of this partition are each $S/2$ for the NN, NNN and parabolic bands with both couplings. Both partitions give the same total $S$ and distribute it differently between the two contributions.

\begin{figure}[tbp]
\centering\includegraphics[width=3.4in]{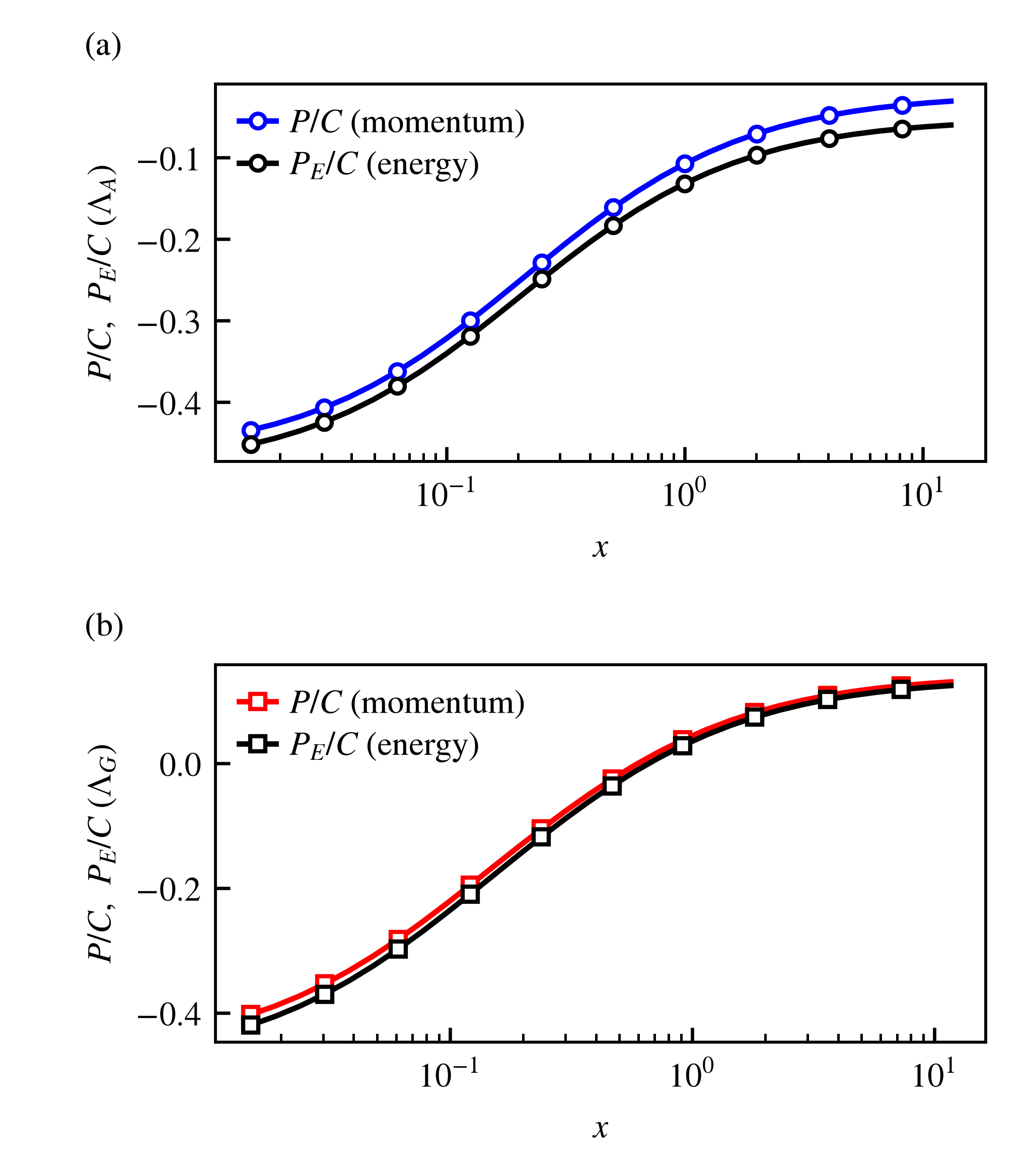}
\caption{Ratio of occupation to charge response in two partitions on the NN reference band along $R=2$, at fixed density, $k_BT/t_1=0.05$ and $\xi\bar k_F=4$. Panels (a) and (b) use $\Lambda_A$ and $\Lambda_G$, respectively. Each panel overlays $P/C$ in the momentum partition (blue for $\Lambda_A$, red for $\Lambda_G$) and $P_E/C$ in the energy partition (black). Open circles and squares distinguish the two couplings. The horizontal axis $x=C_0/C-1$ is the same full charge response in both partitions.}
\label{fig:partitionR2}\end{figure}

We locate the zeros of the NN occupation term on the lines $R=1,2,4,8$ by bisecting each sign change on a grid of 30 logarithmic values of $K$ from $0.01$ to $10$, with $a^*=RK^*/f_m$ from Eq.~\eqref{eq:sm_R}. For $\Lambda_A$, the momentum partition has zeros at $R=4$, at $K^*=4.33$ and a relative magnetic resistivity $x=9.48$, and at $R=8$, at $K^*=0.897$ and $x=3.58$, and the energy partition keeps its sign on this grid. For $\Lambda_G$, both partitions have zeros on all four lines, at relative magnetic resistivities between $0.44$ and $1.04$. Along $R=2$, the zero lies at $K^*=0.463$, $a=11.3$ and $x=0.597$ in the momentum partition, and at $K^*=0.517$, $a=12.6$ and $x=0.663$ in the energy partition. A larger spin flip weight $R$ moves the zero to smaller $K^*$. Figure~\ref{fig:partitionR2} overlays the momentum and energy partitions along $R=2$ for each coupling.

\section{Band survey and the van Hove saddle}\label{sec:bandsurvey}
This section supports Sec.~III~E of the main text. The filling $\nu$ is the occupation per spin per site, with a full band at $\nu=1$, so the total density is $n=2\nu$. The lattice survey uses $t_2/t_1=0,-0.1,-0.2,-0.3,0.1$, $\nu=0.20,0.30,0.40,0.60,0.70$ and $k_BT/t_1=0.02,0.05,0.1$, with $K=0.1$, $\xi\bar k_F=4$, both couplings and $a=1,3,10$. For each density and temperature we solve $\mu(T)$ and evaluate the distance from the saddle in units of temperature, $\delta_{\rm vH}=|\mu-4t_2|/k_BT$, and we use the paired integration of Sec.~\ref{sec:sm_thermal} for $\delta_{\rm vH}<15$. Three cells lie within $2k_BT$ of the saddle and are left out, and the remaining 72 cells give 144 responses for the two couplings at each strength.

\begin{figure}[tbp]
\centering\includegraphics[width=3.4in]{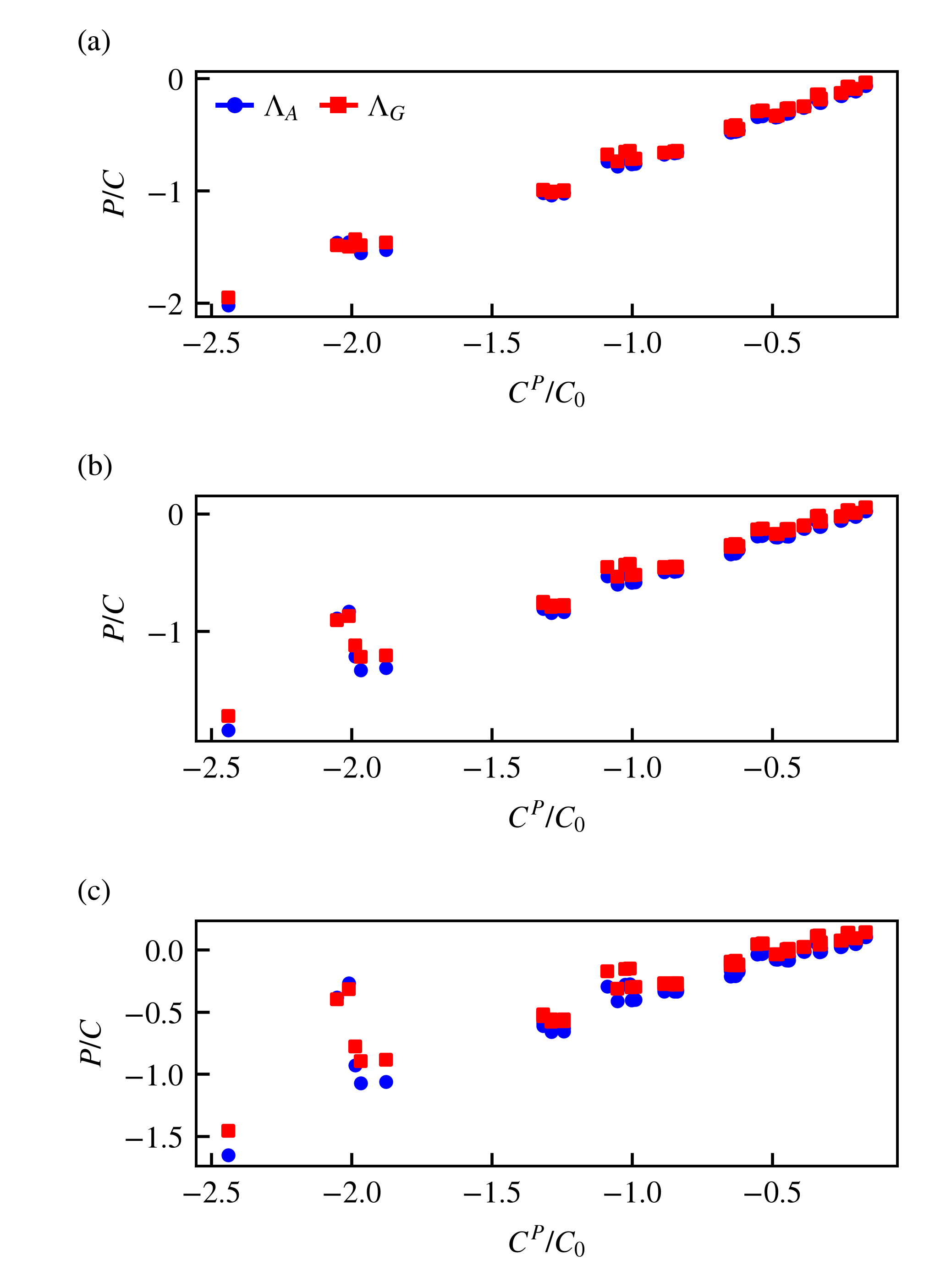}
\caption{Scattered ratio $P/C$ of occupation to charge response against its bare value $C^P/C_0$ in the momentum partition. Panels (a), (b) and (c) show $a=1,3,10$, respectively, at $K=0.1$ and $\xi\bar k_F=4$. Each panel contains the same 72 density and temperature cells, with blue circles for $\Lambda_A$ and red squares for $\Lambda_G$, giving 144 responses. The bare coefficients are evaluated at the same density and temperature with $K=a=0$.}
\label{fig:bandmagnitude}
\end{figure}

Scattering reduces the occupation ratio. Before scattering, the median of $|P/C|$ over the 144 responses is 0.59 and its maximum 2.44. At $K=0.1$ the median falls to 0.38, 0.23 and 0.12 at $a=1$, 3 and 10, and the maximum to 2.02, 1.84 and 1.65. Figure~\ref{fig:bandmagnitude} compares each scattered occupation ratio with its bare value. Scattering shifts $P/C$ upward at every response, with median shifts of $0.17$, $0.32$ and $0.46$ at $a=1$, $3$ and $10$, and the occupation term reverses sign on 12 and 48 responses at $a=3$ and $10$. The parabolic band, with $C^P=0$, acquires a negative occupation term from scattering alone. An extended survey over eight values of $a$ and $K=0.1,0.3$ on 58 cells changes $V/C$ by at most $0.098$, about 12 percent of its bare value, so the relation $V/C=b$ of the NN band holds only for bands with $t_2=0$.

\subsection{Distance from the van Hove saddle}\label{sec:vhdistance}
For every hopping ratio on the grid, the saddles of $\epsilon_0$ lie at $(\pi,0)$ and $(0,\pi)$, where the velocity vanishes and $|d_\kk|=2$ is the maximum of $|\cos k_x-\cos k_y|$ over the zone. By the lattice identity of the main text, $C^P$ is a density of states weighted by $\Phi_d^2$, which equals 4 at both saddles. The neighborhood of the saddles contributes with one sign and with a density of states that grows as $\delta_{\rm vH}$ decreases, while it adds little to $C_0=\int_\kk Fv_x^2$ because the velocity vanishes there. The cell $(t_2/t_1,\nu,k_BT/t_1)=(-0.2,0.40,0.02)$, at $\delta_{\rm vH}=2.09$, gives the largest occupation ratio at every strength, and the responses with $|P/C|>1$ have a median $\delta_{\rm vH}$ of 4.7, against 21 for the rest. For $\delta_{\rm vH}\ge5$, the medians of $|P/C|$ are $0.48$, $0.31$, $0.17$ and $0.09$ at $a=0,1,3,10$. Away from the saddle the occupation term thus remains about one half of the charge coefficient before scattering and one third at $a=1$, while the values above unity belong to bands close to the saddle.

\subsection{Field driven and collision parts across the survey}
The analytic drive is evaluated in the contour representation with the thermal factor included once, and $P_{\rm coll}=S-V-P_{\rm drive}$ follows from the computed $S$ and $V$. On every retained response and at every strength, the field driven part has the sign of $C^P/C_0$. A reversal of the occupation term requires the collision part to oppose and exceed the field driven part. This happens on 0, 12 and 48 responses at $a=1$, $3$ and $10$, of which 0, 3 and 9 lie on bands with $t_2=0$. The energy partition gives 0, 0 and 12 reversals of the same total responses. The change of the collision process with the order is necessary for the lattice reversal, and it dominates the field driven part when spin flip scattering is strong.

\section{Range of linear splitting}\label{sec:sm_windows}
This section supports Sec.~III~E of the main text. We compute the spin splitter ratio $\theta(\phi_0)$ at fixed density and fluctuation strengths and record where $\theta/\theta_{\rm bg}$ first departs from its linear value by 1 percent. We give the first crossing because the dependence is nonmonotonic, and we stop at the first crossing of a spin saddle through the chemical potential, $\phi_{0c}=0.657$ for the NN band and $0.0226$ for the band near the saddle. On the NN band, the full splitting on the Fermi line is $3\phi_0t_1$ and the saddle splitting is $4\phi_0t_1$. The first one percent departure lies at full splittings of $0.32t_1$ to $0.46t_1$ on the NN reference band at $\xi\bar k_F=4$, and of about $0.05t_1$ at $(t_2/t_1,\nu,k_BT/t_1)=(-0.2,0.40,0.02)$ with $K=0.1$.

\subsection{Correlation length and the linear range}
\begin{figure}[tbp]\centering\includegraphics[width=3.4in]{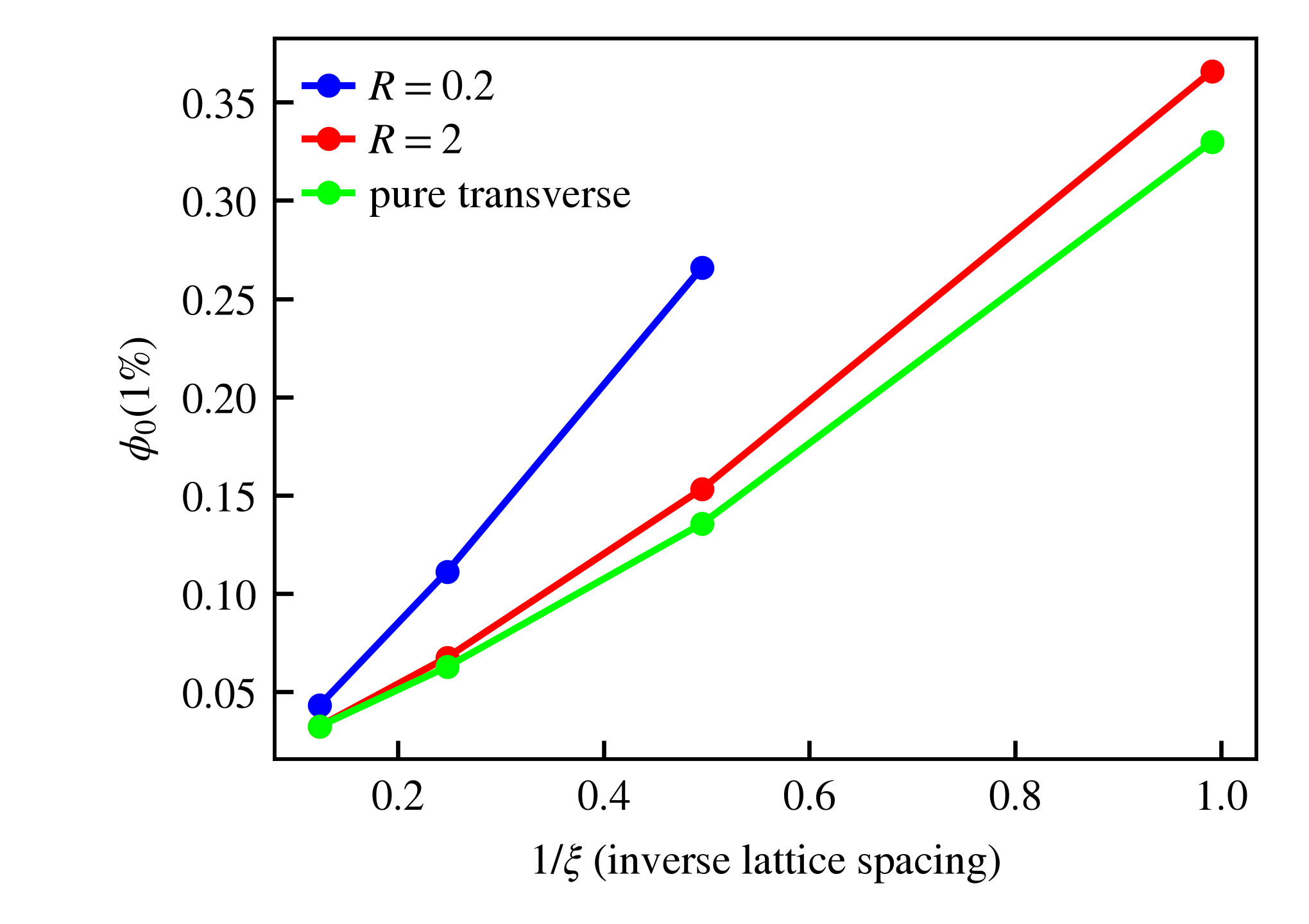}
\caption{Order amplitude at the first 1 percent departure, $|\eta(\phi_0)-1|=0.01$, from the response at linear order on the NN reference band with $\Lambda_A$ and $k_BT/t_1=0.05$, against inverse correlation length in units of the inverse lattice spacing. Blue, red and green circles show $R=0.2$, $R=2$ and purely transverse fluctuations, respectively, all with the same total strength, calibrated as $K(1+R)=0.09$, equal to that of $K=0.03$ and $R=2$ at each $\xi$. Density is fixed. Points are shown where the 1 percent departure lies below $\phi_0=0.519$, the amplitude at which the first spin saddle is $3k_BT$ from the chemical potential.}\label{fig:windowxi}\end{figure}

A spin flip between the split bands requires a momentum transfer of at least $2\phi_0|d_{\mathbf k}|/(\hbar|\mathbf v|)$, so the linear expansion requires $2\phi_0|d_{\mathbf k}|\xi/(\hbar|\mathbf v|)\ll1$. On the NN band at $k_BT/t_1=0.05$, keeping the first spin saddle at least $3k_BT$ from $\mu(\phi_0)$ sets the endpoint $\phi_{0\max}=0.519$. We compare $R=0.2$, $2$ and purely transverse fluctuations at equal total strength, calibrated as $K(1+R)=0.09$ at each correlation length, with the density fixed as the splitting varies. Figure~\ref{fig:windowxi} shows that the first one percent departure moves to smaller $\phi_0$ roughly as $1/\xi$, and the largest momentum mismatch at that departure is about $0.45$ to $0.60$ for $R=2$ over $\xi\bar k_F=2$ to $16$. Figure~5 of the main text and Fig.~3(a) of the companion Letter show the nonlinear reduction for the three baths.

\section{Background of finite range}\label{sec:sm_background}
\begin{figure}[tbp]\centering\includegraphics[width=3.4in]{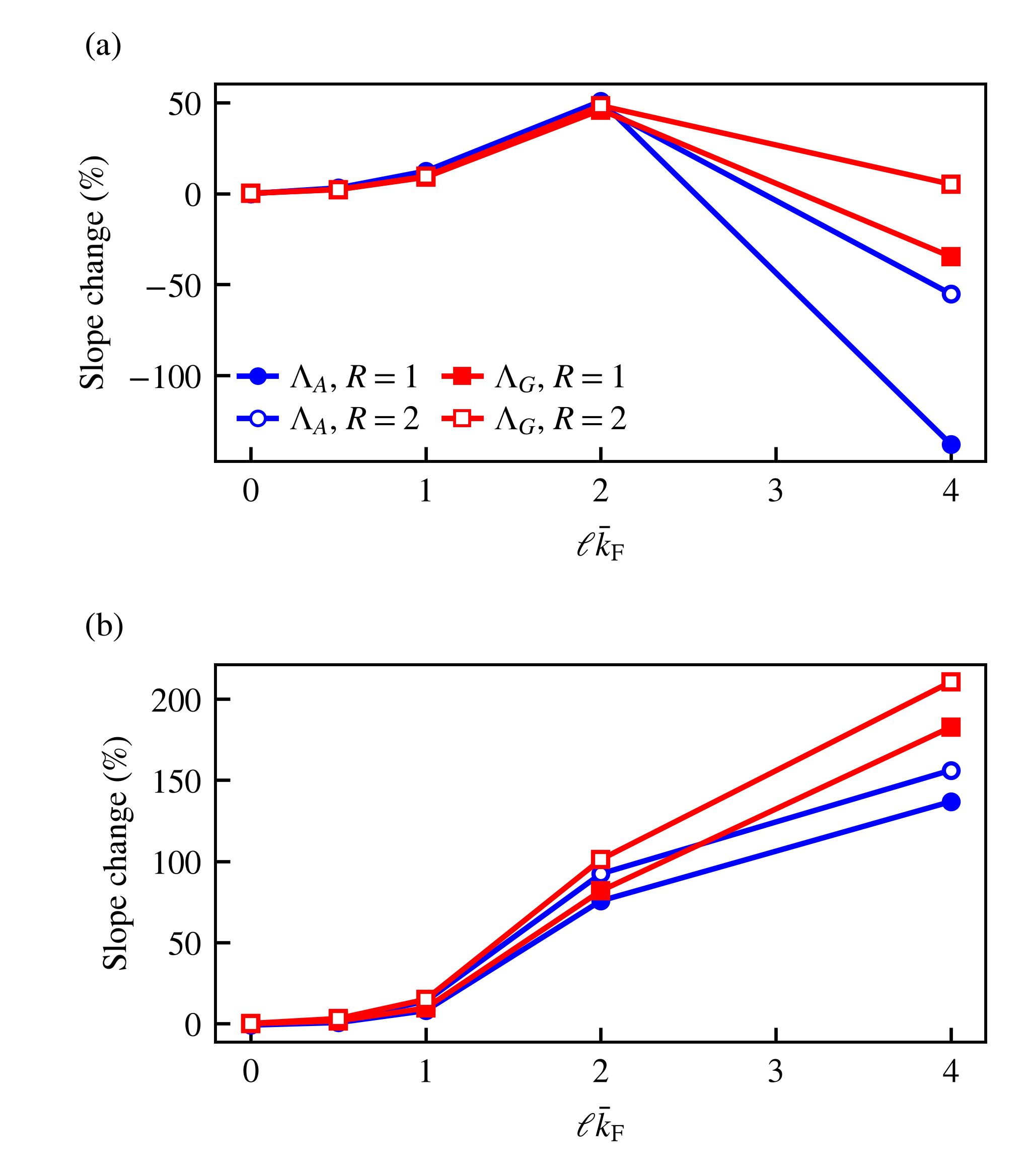}
\caption{Relative change of the slope $s$, $100[s(\ell)/s(0)-1]$, as a function of defect range $\ell\bar k_F$. The reference $s(0)$ uses pointlike spin conserving defects. Panels (a) and (b) use the NN and NNN reference bands, at fixed density, $k_BT/t_1=0.05$ and $\xi\bar k_F=4$. The background transport rate at the unsplit Fermi level is held fixed as $\ell$ changes. Blue circles denote $\Lambda_A$ and red squares $\Lambda_G$. Filled and open markers denote $R=1$ and $R=2$.}\label{fig:finitebg}\end{figure}

This section supports Secs.~III~E and V of the main text. We describe spin conserving defects of range $\ell$ by a scattering probability proportional to $\exp(-q_{\rm lat}^2\ell^2/2)$, normalized so that the background transport rate at the unsplit Fermi level stays fixed. The scattering amplitude stays fixed on all energy and split contours, and scattering into a state and its change with the order are kept. For pointlike defects, $\ell=0$, the reference reduced spin splitter ratio is $Q=-0.965$ on the NN band and $-1.68$ on the NNN band, and the slopes change by less than one percent compared with a constant background. At $\ell\bar k_F=1$ the reference values shift by 5.7 and 3.5 percent and the slopes by 8 to 15 percent. At $\ell\bar k_F=2$ the slopes change by about 50 percent on the NN band and 75 to 100 percent on the NNN band, and at $\ell\bar k_F=4$ the NN reference value shifts by about $-60$ percent and one NN slope becomes negative. A background of finite range relaxes spin and charge currents at different rates. Varying the defect concentration at fixed range still defines a disorder series, with the slope that belongs to that background. Figure~\ref{fig:finitebg} resolves the change of slope for each coupling and $R$.

\subsection{An anisotropic background in a disorder series}
Electron-phonon scattering can change the spin-splitter response \cite{SMHe2025}.
To estimate the phonon term of the disorder series in Sec.~V of the main text, we combine pointlike defects with spin conserving scattering of range $\ell\bar k_F=2$, which serves as a model of anisotropic scattering. The anisotropic scattering has a fixed absolute strength, while the concentration of pointlike defects varies. With magnetic scattering switched off, its slope is $s_{\rm ph}=-0.146$ on the NN band and $-0.0468$ on the NNN band.

\begin{figure}[tbp]\centering\includegraphics[width=3.4in]{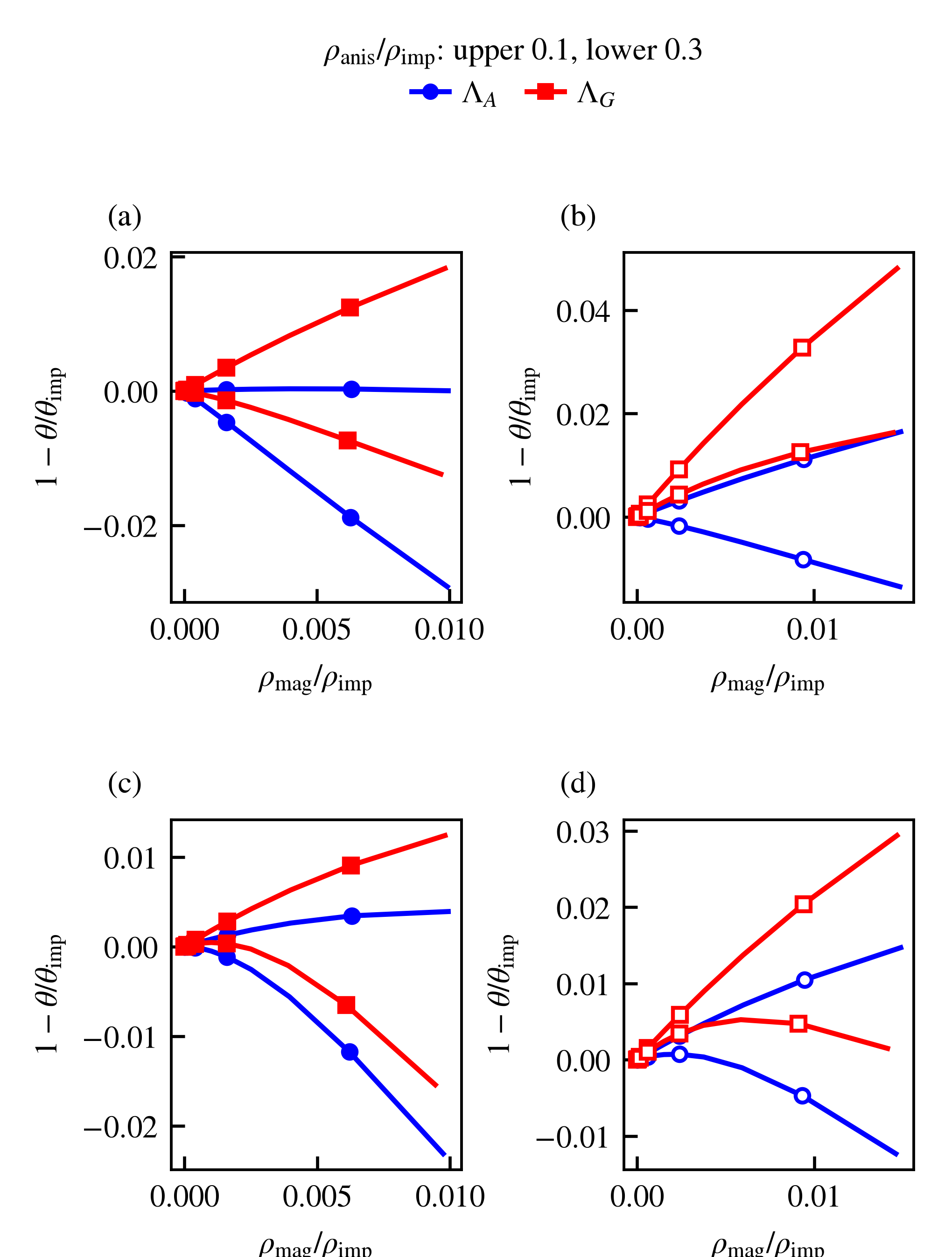}
\caption{Disorder series with fixed magnetic scattering and a fixed spin conserving background of range $\ell\bar k_F=2$, while the concentration of pointlike defects varies. The ordinate is $1-\theta/\theta_{\rm imp}$, with $\theta_{\rm imp}$ the reference with point defects alone, and the abscissa is $\rho_{\rm mag}/\rho_{\rm imp}$. The upper and lower rows use the NN and NNN reference bands, at fixed density, $k_BT/t_1=0.05$ and $\xi\bar k_F=4$. The left and right columns have $R=1$ and $R=2$. Blue circles denote $\Lambda_A$ and red squares $\Lambda_G$. For each coupling, the upper and lower curves correspond to ratios 0.1 and 0.3 of anisotropic to defect resistivity in the reference sample with $K=0.01$.}\label{fig:anisbg}\end{figure}

Figure~\ref{fig:anisbg} follows magnetic scattering with $R=1,2$ and $K=0.01$ in a reference sample whose ratio of anisotropic to defect resistivity is 0.1 or 0.3. That ratio changes along the disorder series because only the defect concentration varies, while the fixed anisotropic and magnetic strengths keep $\rho_{\rm ph}/\rho_{\rm mag}$ constant. A slope accurate to 10 percent requires $\rho_{\rm ph}/\rho_{\rm mag}<0.1|s|/|s_{\rm ph}|$, which gives bounds of 1.12 to 3.42 on the NN band and 2.91 to 6.10 on the NNN band, depending on the coupling and $R$, independent of the defect concentration.

\begingroup
\let\SMoriginalLabel\label
\renewcommand{\label}[1]{\SMoriginalLabel{SM:#1}}

\endgroup

\end{document}